\documentclass[aps,prx,twocolumn,amsmath,amssymb,showpacs,superscriptaddress,notitlepage,longbibliography,floatfix]{revtex4-2}
\usepackage{graphicx}
\usepackage{subfigure}
\usepackage{dcolumn}
\usepackage{bm}
\usepackage{amsfonts}
\usepackage{mathrsfs}
\usepackage{amssymb}
\usepackage{amsmath}
\usepackage{color}
\usepackage{braket}
\usepackage{chngcntr}
\usepackage{stmaryrd}
\usepackage{ulem}
\usepackage[colorlinks=true,breaklinks=true,linkcolor=red,anchorcolor=red,citecolor=red,urlcolor=red]{hyperref}
\usepackage{booktabs}
\usepackage{multirow}
\usepackage{txfonts}
\usepackage[T1]{fontenc}

\begin{document}

\title{Non-Hermitian Topology from Edge Transport in Hermitian Quantum Anomalous Hall Systems}
	\date{\today}
	
\author{Humian Zhou}
\affiliation{International Center for Quantum Materials, School of Physics, Peking University, Beijing 100871, China}
\author{Ming Lu}
\affiliation{Beijing Academy of Quantum Information Sciences, Beijing 100193, China}
\author{Chui-Zhen Chen}
\email{czchen@suda.edu.cn}
\affiliation{School of Physical Science and Technology, Soochow University, Suzhou 215006, China}
\affiliation{Institute for Advanced Study, Soochow University, Suzhou 215006, China}
\author{X. C. Xie}
\email{xcxie@pku.edu.cn}
\affiliation{International Center for Quantum Materials, School of Physics, Peking University, Beijing 100871, China}
\affiliation{Institute for Nanoelectronic Devices and Quantum Computing, Fudan University, Shanghai 200433, China}
\affiliation{Hefei National Laboratory, Hefei 230088, China}

\begin{abstract}
	Non-Hermitian physics, known for exotic phenomena like exceptional points and the skin effect, has 
    been most prominently realized in
    engineered systems relying on controlled gain and loss. Here we show that it can also arise naturally as an intrinsic transport response of a globally Hermitian quantum anomalous Hall system, without the need for external non-Hermitian engineering.	 
    We show that the interplay between unidirectional chiral edge modes and diffusive normal edge modes induces intrinsic non-reciprocal transport described by a  continuum Hatano-Nelson model.   
    Consequently, the non-Hermitian skin effect is encoded directly in experimentally accessible Hall-bar observables: the electrochemical potential and local heat dissipation acquire chirality-dependent exponential spatial profiles, while the longitudinal conductance decays exponentially with system size and the Hall conductance remains quantized.    
    Using Landauer--B{\"u}ttiker simulations,
    we confirm these transport signatures and identify magnetic topological insulators as a realistic platform for an intrinsic non-Hermitian transport response. Our results bridge non-Hermitian topology with mesoscopic transport, opening a pathway toward non-Hermitian topological devices in solid-state systems.
\end{abstract}

	\maketitle

Non-Hermitian (NH) physics has emerged as a fundamental paradigm for open systems coupled to external environments \cite{Bender1998,Heiss2012,Bergholtz2021}.
The discovery of exceptional points, exotic Non-Hermitian topological phases, and the non-Hermitian skin effect (NHSE) has drastically enriched the landscape of quantum mechanics beyond the conventional Hermitian framework \cite{Lee2016,Kunst2018,Yao2018,xiong2018,Martinez2018,Gong2018,Kawabata2019,Shen2018,Yokomizo2019,Zirnstein2021,Zirnstein2021b,Yao2018b,Borgnia2020,Xue2021,Yi2020,Kornich2023,Budich2020,Wang2022}. 
To date, however, experimental realizations of these Non-Hermitian phenomena have predominantly relied on engineered platforms with externally imposed gain and loss \cite{Zhao2025,Liang2022,Xiao2020NHBBC,Wang_2021,Helbig2020BBC,Liu2021NHSEcircuit,Sebastian2020,Brandenbourger2019,Ghatak2020,Zhang2021a,Zhang2021b}.
Meanwhile, recent studies have shown that NH topology can also be encoded in Hermitian topological matter \cite{Selma2022,Shu2024,Daichi2025,Okuma2021}, suggesting that NH physics may also be explored directly through transport phenomena in Hermitian solid-state systems.

Magnetic topological insulators (TIs) hosting the quantum anomalous Hall (QAH) state with chiral edge modes and the axion insulator state \cite{Chang2023,Haldane1988,Liu2008,Yu2010,Chang2013,Kou2014,Chang2015,Charles2020,LiHaiLong2021,okazaki2022quantum,Jiang2012,Wang2013,chang2015high,XiaoDi2018,chen2020tunable,Yujun2020, checkelsky2014trajectory,Chen2019,chen2019intrinsic,liu2020robust,Lian2025}  provide a particularly promising solid-state  platform for exploring NH phenomena.  Previous studies have approached NH physics in Hermitian topological electronic systems in several ways \cite{Selma2022,Shu2024,Daichi2025,Ochkan2024,Fulga2025,Le2026,ochkan2025,Selma2022,Shu2024,Daichi2025}. In some cases, disorder or interactions generate effective self-energies associated with finite quasiparticle lifetimes, leading to NH band structures with momentum-space topology \cite{Kozii2024,Papaj2019}. In others, NH topology is encoded in reflection matrices or in effective boundary Hamiltonians obtained by integrating out the bulk \cite{Selma2022,Shu2024,Daichi2025}. More recently, quantum Hall and QAH devices have provided a route to NH topology through multi-terminal conductance matrices, in which  contact arms act as effective sites of a discrete Hatano-Nelson chain \cite{Ochkan2024,Fulga2025,Le2026,ochkan2025}.
Despite these important developments, directly isolating transport signatures of NH topology in a globally Hermitian electronic systems, particularly the chirality-dependent exponential transport response associated with the NHSE, remains challenging.
This motivates our central inquiry: can NH topology arise directly from the intrinsic edge dynamics of a globally Hermitian QAH system, with signatures appearing naturally in the spatially continuous evolution of local, experimentally measurable transport observables?

\begin{figure}[bht]
\centering
\includegraphics[width=3.3in]{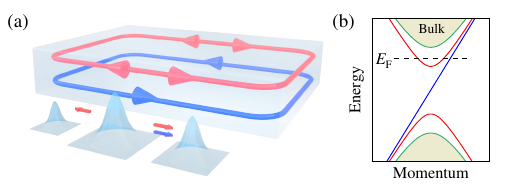}
\caption{ 
	Nonreciprocal transport and NHSE in QAH edge states. (a) Schematic plot of unidirectional chiral (blue) and bidirectional normal (red) edge modes. The nonreciprocal transport is highlighted by the unequal amplitude splitting of edge-injected wave packets.
	This imbalance serves as a dynamic signature of the NHSE.
	(b) Corresponding energy-momentum dispersion, showing the chiral (blue) and normal (red) edge modes  relative to the Fermi level $E_F$.
	\label{fig3}}
\end{figure}

In this work, 
we show that realistic QAH insulators provide a solid-state transport setting in which non-Hermitian topology emerges directly from intrinsic edge dynamics.
While ideal QAH insulators host only chiral edge modes, realistic samples often harbor non-topological normal edge modes due to imperfections \cite{Chang2015,Charles2020,Chang2023}. We show that the interplay between a chiral mode and a normal edge mode realizes the continuum Hatano-Nelson model.
By mapping the edge transport equations to the HN  Hamiltonian,  we demonstrate that a Hermitian QAH system exhibits characteristic signatures of Non-Hermitian topology, including the directional exponential behavior of the Green's function and the NHSE.
Here, the term ``Hermitian QAH system'' refers to the QAH sample described by a Hermitian microscopic Hamiltonian.
We further show that these features manifest in the experimentally measurable transport quantities of a Hall-bar device, where heat dissipation and the longitudinal conductivity acquire an exponential behavior while the Hall response remains robustly quantized. Our results uncover a condensed-matter realization of non-Hermiticity rooted in microscopic chiral dynamics, bridging Non-Hermitian topology with mesoscopic transport.

\par
\vspace{1em}
\noindent \textbf{\large{Quantum Kinetic Theory of QAH Edge States}}\\
We begin by investigating the microscopic theory of edge transport in a 2D QAH insulator. 
As illustrated in Fig.~\ref{fig3}, we consider the QAH insulator that hosts both a chiral edge mode and a normal edge mode. The coexistence of these modes has been confirmed by transport experiments on magnetic TI samples \cite{Kou2014,Chang2015,Charles2020}. 
The total Hamiltonian of the edge system is given by:
\begin{align} \label{Ham}
	H = H_0 + U(x) + U_D(x),
\end{align}
where $H_0=\mathrm{diag}\{\varepsilon_{1k},\varepsilon_{2k}\}$ describes the normal  and chiral   modes with dispersions $\varepsilon_{1,k}=\varepsilon_{1,-k}$ and $\varepsilon_{2k} = cv_c \hbar k$, respectively. The parameter $v_c$ denotes the chiral mode velocity, and $c=\pm 1$ indicates the propagation direction.
The term  $U(x)$ represents a slowly varying electrostatic potential, while $U_D(x)=\sum_{\alpha,  i}u_{\alpha }\sigma_{\alpha} \delta(x-R_{i})$  denotes the delta-function impurity potential with the randomly distributed impurity position $R_{i}$. Here, the Pauli matrices \( \sigma_{\alpha} \) ($\alpha = 0,x$) act on the band indices. The parameters $u_{0}$ and $u_{x}$ represent the intra-band and inter-band scattering strengths, respectively.

To model the transport, we utilize the quantum Liouville equation for the density operator $\hat{\rho}_t$:
\begin{align} \label{QLE}
	\frac{\partial \hat{\rho}_t}{\partial t}+\frac{i}{\hbar}[\hat{H},\hat{\rho}_t]=0.
\end{align}
After performing disorder averaging under the first-order Born approximation and applying the Wigner transformation(see Supplemental Material \cite{SM} for details), we obtain the semiclassical Boltzmann transport equation:
\begin{align}
	&\frac{\partial f_{nk}}{\partial t} 
	+ v_{nk} \partial_x f_{nk}
	- \partial_x U \frac{\partial f_{nk}}{\hbar \partial k} \nonumber \\
	&= - \sum_{n'} \int \frac{\mathrm{d}k'}{2\pi  }\,
	W_{nk,n'k'} \big( f_{nk} - f_{n'k'} \big)
\end{align}
Here, $f_{nk}$ is the distribution function for state $k$ in band $n$ and
$v_{nk} =\hbar^{-1}\partial_k \varepsilon_{nk}$ is the group  velocity.
The scattering rate is defined as
$W_{nk,n'k'}={2\pi}{\hbar}^{-1}\sum_\alpha n_\alpha u_\alpha^2 
(\sigma_{\alpha})_{{nn'}} (\sigma_{\alpha})_{{n'n}}
\delta(\varepsilon_{nk} - \varepsilon_{n'k'} )$, where $n_{\alpha}$ is the impurity density. 

We consider a weakly nonequilibrium state induced by a slowly varying electrochemical potential $\tilde{\mu}(x)$.
The small statistical force $\partial_x \tilde{\mu}$ produces only weak deviations of $f_{nk}$ from  equilibrium state. 
Thus, we linearize the non-equilibrium distribution as
$f_{nk} \simeq  f_0(\varepsilon_{nk} - \mu) + f_{nk}^{(1)}$.
Here $f_0(\varepsilon_{n k}-\mu)$ denotes the Fermi--Dirac distribution, and $\mu(x)=\tilde{\mu}(x)-U(x)$ is the local chemical potential that determines the local equilibrium occupation.
The first-order corrections are derived as: \(
f^{(1)}_{1k} = [\mathrm{sign}(k)v_{1F} \tau_1 -   c v_{2F} \tau_2g_2/g_1]
{\partial_{\varepsilon} f_0(\varepsilon_{1k}-\mu)}\, \partial_x \tilde{\mu}\), and 
\(
f^{(1)}_{2k} = cv_{2F}\tau_2
{\partial_{\varepsilon} f_0(\varepsilon_{2k}-\mu)} \, \partial_x \tilde{\mu}.
\)
Here, $\tau_1$ and $\tau_2$ are the relaxation time for the distributions of normal edge channel ($f_{1k}$) and the chiral edge channel ($f_{2k}$) to reach equilibrium \cite{SM}, while  $v_{1F}=|\hbar^{-1}\partial_k \varepsilon_{1k}|_{\varepsilon_{1k}=E_F}$ and  $v_{2F}=v_c$  correspond to Fermi velocities. $E_F$ is the Fermi level.
The explicit forms of the relaxation time are ${\hbar/\tau_1}= 
{2\pi}(n_0u_0^2g_1+n_xu_x^2g_2)$
and $\hbar/{\tau_2} ={2\pi}n_xu_x^2
(g_1+g_2)^2/g_1$,
where $g_n = \int \! \frac{\mathrm{d}k}{2\pi} \, \delta(E_F - \varepsilon_{nk})$ is the density of states (DoS) of band $n$ at Fermi energy.
Here, the transport lifetime for the normal edge channel ($\tau_1$) is governed by both scattering strengths $u_{0}$ and $u_{x}$.
Conversely, the relaxation time of the chiral edge channel ($\tau_2$) is determined solely by the inter-band scattering strength $u_x$. These results demonstrate that, while the chiral edge state is immune to intra-band backscattering (thereby preserving its unidirectional nature in isolation), the coexistence of a normal mode introduces dissipation via inter-band scattering.

The edge current is given by $j=-e\sum_n \int \! \frac{\mathrm{d}k}{2\pi} \, v_{nk} f_{nk}$, and takes the form
\begin{eqnarray}\label{eq_j}
	j(x) = \frac{1}{e}\sigma_D \partial_x \tilde{\mu}(x) - c\frac{e}{h}(\tilde{\mu}(x)-\tilde{\mu}_0) + j_0.
\end{eqnarray} 
Here, $\sigma_D = e^2 (g_1  D_1 + g_2 D_2 )$ is the one-dimensional conductivity with diffusion constant $D_n= (v_{nF})^2 \tau_n$, 
which arises from backscattering due to the normal edge modes. 
$\tilde{\mu}_0$ denotes the constant electrochemical potential at equilibrium.
The first term in Eq.~\ref{eq_j} corresponds to the diffusive current driven by the statistical force $\partial_x \tilde{\mu}$. The second term describes the chiral current carried by the chiral edge mode.
The final term, 
$j_0$, represents a circulating equilibrium current, which does not contribute to charge transport.
Eq.~\ref{eq_j} remains unchanged even when the inelastic scattering is included, whose main effect is to renormalize  $\sigma_D$ \cite{SM}.

The charge continuity equation is given by $\partial \rho /\partial t  + \partial_x(j-j_0) = S$, where $S$ is the source term and the charge density is given by $\rho = -e\sum_n \int \! \frac{\mathrm{d}k}{2\pi} \, f_{nk}=\rho_0 -e g_0(\tilde{\mu}-\tilde{\mu}_0) + \mathcal{O}((\tilde{\mu}-\tilde{\mu}_0)^2)$ at low temperature. $-eg_0=\partial \rho/\partial \tilde{\mu} $ and $\rho_0$ is the charge density at equilibrium.  In the absence of the source term, we obtain the following effective Schr{\"o}dinger-like equation of motion: 
\begin{eqnarray}\label{eq_rho}
	i\hbar \frac{\partial \tilde{\mu}}{\partial t}  = -i[\frac{{\hat{p}^2}}{2m} + icv\hat{p}]\tilde{\mu} \equiv -iH_{\text{HN}}(\hat{p})\tilde{\mu}.
\end{eqnarray}
Here, $\hat{p}=-i\hbar\partial_x$ is the momentum operator, $m=e^2 g_0\hbar/(2\sigma_D)$ is the effective mass, and $v=1/(h g_0)$ denotes the drift velocity.
Equation~\eqref{eq_rho} demonstrates that edge transport in a QAH insulator is governed by the Non-Hermitian Hamiltonian $H_{\text{HN}}$, 
which represents the continuum limit of the HN model \cite{Hatano1996} derived in the long-wavelength limit \cite{SM}.
 Notably, the Non-Hermitian topology of this system stems directly from the inherently nonreciprocal propagation of the chiral edge modes.
Unlike the wavefunction in the standard Non-Hermitian Schr{\"o}dinger equation, the electrochemical potential $\tilde{\mu}$ is a directly measurable quantity in experiments. This distinction opens up a new avenue to investigate novel effects in Non-Hermitian systems.
\begin{figure}[bht]
	\centering
	\includegraphics[width=3.3in]{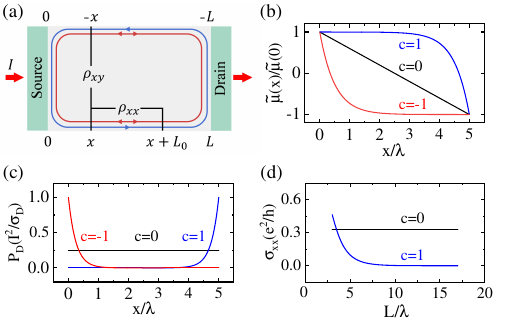}
	\caption{ 
		(a) Schematic of a Hall-bar device hosting one chiral and one normal mode. The positive directions of the local 1D boundary coordinates for the upper and lower edges follow their respective chiral propagation directions.
		(b),(c) Spatial profiles of the electrochemical potential (b) and the heat dissipation (c) along the lower edge of a QAH sample for $c=0,\pm 1$ with $L=5\lambda$.
		(d) The longitudinal $\sigma_{xx}$ as a function of sample length $L$ for $c = 0 $ and 1 with $x = 0.4L$, $L_0=0.3L$, $W =3\lambda$ fixed.
		\label{fig1} }
\end{figure}
\par
\vspace{1em}
\noindent \textbf{\large{NHSE of Non-Hermitian Edge Dynamics}}\\
To explore the NHSE in the Non-Hermitian Hamiltonian, we investigate the Green's function of the system under a periodic boundary condition (PBC), defined as $G(E,x,x') = \bra{x}[E-H_{\text{HN}}]^{-1}\ket{x'}$.
It takes the form \cite{SM}:
\begin{eqnarray}\label{eq_Green}
	G(E;x,x')  = -\frac{im}{\hbar^2\kappa}
	\left(
	\frac{e^{i \kappa (x-x')}}{1 - e^{i 2L k_+}}-
	\frac{e^{-i \kappa (x-x')}}{1 - e^{i 2L k_-}}
	\right)e^{\frac{c(x-x')}{\lambda}}
\end{eqnarray}
which is valid for the interval $0\leq x-x'<2L$. For other coordinate differences, the Green's function satisfies the periodicity condition $G(E;x,x'\pm 2L) = G(E;x,x')$. 
$L$ is the length of the QAH sample.
Here, $\kappa = \sqrt{2mE-(cmv)^2}/\hbar$ and $k_{\pm}= -ic/\lambda \pm \kappa$ with $H_{\text{HN}}(\hbar k_{\pm})=E$.
The parameter $\lambda = \hbar/(mv)=2\sigma_D h/e^2$ serves as a characteristic length scale governing the exponential behavior of the Non-Hermitian system.  $\lambda$ is determined solely by the normal-mode diffusive conductivity $\sigma_D$. This implies that stronger impurity scattering (or a reduced number of normal modes) yields a smaller $\sigma_D$ resulting in a shorter $\lambda$ and thus a faster exponential decay.

Remarkably, the Green's function exhibits a distinct directional exponential amplification that becomes fully visible under open boundary conditions (OBC). In this case, the right eigenstates of $H_{\text{HN}}$, $\psi_{n,R}(x) = L^{-1/2} e^{c x / \lambda} \sin(n\pi x/2L)$ with $n \in \mathbb{Z}$
, are exponentially localized at the boundary for $c=\pm1$, directly signaling the NHSE. 
The NHSE is the manifestation of the unique Non-Hermitian topology\cite{Zhang2020,Okuma2020}.
The Non-Hermitian topological winding number is defined as  $w(E)= \frac{1}{2\pi i}\int \mathrm{d}k \frac{\mathrm{d}}{\mathrm{d}k}\mathrm{ln}[H_{\mathrm{HN}}(\hbar k)-E]$. For our model, $w(E) = \frac{1}{2}[ \mathrm{sign}(\mathrm{Im}(k_+)) + \mathrm{sign}(\mathrm{Im}(k_-)) ]$, which can take nonzero values for $c=\pm 1$, leading to the emergence of the NHSE\cite{Zhang2020,Okuma2020}.
In contrast, $w(E)$ is always zero for $c=0$, corresponding to a topologically trivial phase without the NHSE.
The correspondence between the exponential spatial asymmetry of the Green's function under PBC and the eigenstate localization under OBC constitutes a hallmark of Non-Hermitian topology. This result explicitly demonstrates the presence of the NHSE in our model. 

\begin{figure*}[bht]
	\centering
	\includegraphics[width=7in]{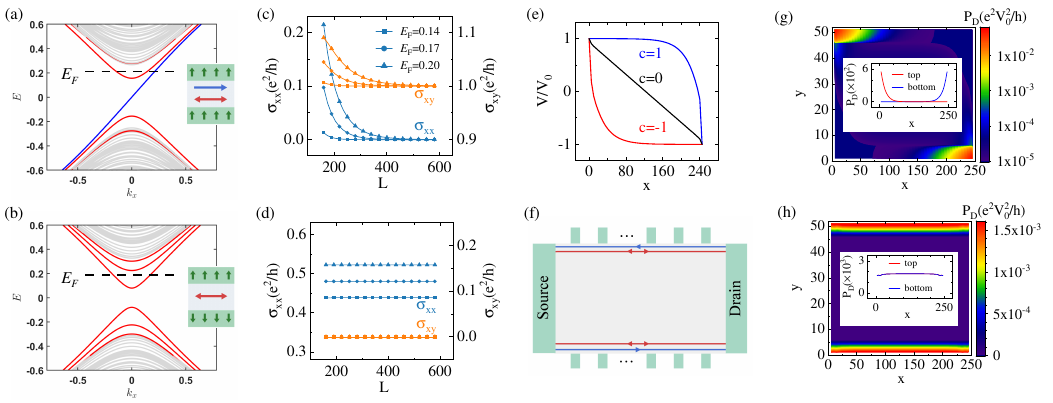}
	\caption{ 
		(a),(b) Energy spectra of the surface states for parallel ($c=1$,QAH) and antiparallel ($c=0$, Axion insulator) magnetizations, respectively. 
		The coexistence of chiral (blue) and normal (red) edge states alongside bulk states (gray) reproduces the band dispersions shown in Fig.~4o of Chang et al. \cite{Chang2015}. 
		Insets: Schematic of the corresponding magnetization and edge-state configurations.		
		(c),(d) Longitudinal ($\sigma_{xx}$) and  Hall ($\sigma_{xy}$) conductances versus system size $L$ for (c) $c =1$ and (d) $c =0$ at various Fermi energies $E_F$.
		(e) Spatial distribution of voltage along the QAH sample edge  for $c=0, \pm1$ with $E_F=0.2$, measured using the multi-terminal Hall-bar geometry illustrated in (f).
		(g),(h) Spatial distribution of the local heat dissipation $P_D$ for (g) $c =1$, (h) $c =0$ with $E_F=0.2$.
		The model parameters are fixed as $A_{x,y}=1.0$, $A_{z}=0.4$, $B_{x,y}=0.4$, $B_z=0.5$, $M_0=1.0$, $M_t=0.4$, and $M_b=0.4$ ($M_b=-0.4$) for QAH (Axion insulator) state. $L=240$ in (e-h).
		Other parameters are $x = L/3$, $L_0 = L/3$, $W = 100$, and the sample thickness $L_z = 6$. Here, we set the lattice constant to unity.
		\label{fig2} }
\end{figure*}

This exotic exponential behavior can be observed on the edge transport of a QAH insulator. We consider a setup with a source at $x=0$ and a drain at $x=L$ [see Fig.~\ref{fig1}(a)], described by the source term $S(x) = I\delta(x) - I\delta(x-L)$.
Solving the steady-state charge continuity equation yields the following relation for the electrochemical potential:
\begin{eqnarray}\label{eq_V}
	\tilde{\mu}(x) = \frac{\hbar I}{eg_0}[G(0;x,0)-G(0;x,L)]
\end{eqnarray}
This equation directly links the experimentally measurable electrochemical potential (or voltage profile, $V(x)=-\tilde{\mu}(x)/e$) to the Green's function of the Non-Hermitian Hamiltonian.
In addition, the local heat dissipation $P_D(x)$ provides another experimentally accessible quantity  \cite{Marguerite2019,Dorri2017,Fang2021,Hailong2024,Qing2024} as a probe of NHSE. 
In our model, $P_D(x) = {\sigma_D}/{ e^2}(\partial_x\tilde{\mu})^2$, which reduces to 
$P_D(x) =  (I^2/\sigma_D) e^{4cx/\lambda}/(1+e^{2cL/\lambda})^2$ for $0\leq x\leq L$ \cite{SM}. 
Consequently, when $c=\pm1$, $\tilde{\mu}$ and $P_D$ exhibit a distinct spatial exponential profile along the lower edge [see Fig.~\ref{fig1}(b) and (c)], whereas for $c=0$, this exponential behavior disappears. 
For the upper edge ($-L\leq x\leq0$), the exponential spatial profile also appears, but the localization occurs at the opposite end of the Hall bar \cite{SM}.
Thus, 
observing these exotic features in the transport of a QAH insulator	can provide clear experimental evidence of the robust NHSE.

Experimentally, the longitudinal and Hall resistance are typically measured using the Hall-bar device of size $L \times W$ (see Fig.~\ref{fig1}(a)). The longitudinal and Hall resistances are obtained from the voltage measurements as $\rho_{xx} = [V(x)-V(x+L_0)]W/(IL_0)$ and $\rho_{xy} = [V(x)-V(-x)]/I$. The corresponding longitudinal and Hall conductance are obtained from the tensor relations: $\sigma_{xx}=\rho_{xx}/(\rho_{xx}^2+\rho_{xy}^2)$, and $\sigma_{xy}=\rho_{xy}/(\rho_{xx}^2+\rho_{xy}^2)$.
Based on this, the longitudinal conductance is expressed as:
\begin{equation}\label{eq_s}
	\sigma_{xx} =
	\begin{cases}
		\dfrac{2\sigma_D}{W}=\dfrac{e^2}{h}\dfrac{\lambda}{W},
		~ \mathrm{for} ~ L\ll\lambda ~ \mathrm{or} ~ c=0\\[6pt]
		\\
		\begin{aligned}
			|c|\,\dfrac{e^2}{h}\,
			\dfrac{W}{L_0}
			\Bigl[
			&\Theta(c)\, e^{2c(x-L_0 - L)/\lambda} \\
			&+\Theta(-c)\, e^{2c x/\lambda}
			\Bigr],
		\end{aligned}
		~\mathrm{for}~ x,L \gg \lambda .
	\end{cases}\nonumber
\end{equation}
where $\Theta(c)$ is the Heaviside step function.
This equation describes $\sigma_{xx}$ in two distinct regimes as shown in Fig.~\ref{fig1}(d).
For $L \ll \lambda$ or $c = 0$, transport is dominated by the normal edge modes, 
and $\sigma_{xx}$ exhibits a classical ohmic law. 
In contrast, for $L \gg \lambda$, the chiral edge mode begins to influence electric transport, 
leading to an exponential dependence of $\sigma_{xx}$ either on $x$ at fixed $L$ or on $L$ at fixed $x/L$.
Therefore, $\sigma_{xx}(c=1)\ll \sigma_{xx}(c=0)$ for $L\gg \lambda$, providing  direct experimental evidence for the  exponential decay of conductance in QAH state.
Such a strong contrast in longitudinal conductance between QAH state ($c=1$) and axion insulator state ($c=0$) has already been observed in magnetic topological insulators (TIs) experimentally \cite{Mogi2017,XiaoDi2018,liu2020robust}.
Previously, exponential decaying of conductance was observed mostly in systems with Anderson localization\cite{Anderson1958, Lee1985} or quantum tunneling\cite{razavy2013quantum}.
Our results show that Non-Hermitian topology offers an alternative mechanism for such exponential behavior in condensed-matter systems.
\\
\\
\noindent \textbf{\large{Experimental Realization in Magnetic TIs}}\\
To demonstrate the experimental feasibility of our proposal, we investigate the transport properties of edge states in magnetic TIs using the Landauer-B{\"u}ttiker formalism \cite{SM,Buttiker1986,Buttiker1988,Yanxia2008}.
We consider a 3D TI with independent magnetization on the top and bottom surfaces (see the insets of Figs.~\ref{fig2}(a-b)). 
The system is described by the four-band effective Hamiltonian \( H_{\text{MTI}} = H_{\text{TI}} + H_M \), where $H_{\text{TI}}(\mathbf{k}) = \sum_{i=x,y,z} A_i k_i \sigma_x \otimes s_i + (M_0 - \sum_{i=x,y,z}B_i k_i^2) \sigma_z \otimes s_0$
represents a 3D TI \cite{Zhang2009} with model parameters \( A_i \), \( B_i \), and \( M_0 \).  
Here, \( \sigma_i \) and \( s_i \) denote Pauli matrices acting on the orbital and spin degrees of freedom, respectively. The Zeeman splitting term is  
\( H_M = M(z)\sigma_0 \otimes s_z \), 
where \(M(z) = M_t\) on the top surface, \(M(z) = M_b\) on the bottom surface, and \(M(z) = 0\) elsewhere. 

Figures.~\ref{fig2} (a) and (b) illustrate the band structures of TI under different magnetization configurations.
For parallel magnetization, the magnetic TI realizes a QAH state with Chern number $c=\pm 1$, characterized by the emergence of chiral edge states traversing the bulk energy gap and 1D normal edge states residing on the side surfaces.
Upon magnetic reversal to an antiparallel alignment induced by an external magnetic field,
the system transitions into an axion insulator with $c=0$. In this regime, the Hall conductance vanishes, and only the 1D normal states persist.
Crucially, the transport in these edge states is governed by the Non-Hermitian HN Hamiltonian $H_{\text{HN}}$, as derived in Eq.~\ref{eq_rho}.

Figures~\ref{fig2}(c) and \ref{fig2}(d) show $\sigma_{xy}$ and $\sigma_{xx}$ as functions of the system length $L$ for different magnetization configurations.
For the QAH state ($c=1$), $\sigma_{xy}$ rapidly approaches the quantized value $e^2/h$, while $\sigma_{xx}$ decreases toward zero with increasing $L$. This confirms that the quantization of $\sigma_{xy}$ remains robust even in the presence of normal edge modes.
In contrast, for the axion insulator state ($c=0$), $\sigma_{xy}$ vanishes and $\sigma_{xx}$ is independent of the system size, consistent with transport through only normal edge modes.
Notably, the NHSE is directly manifested in the exponential voltage distribution $V(x)/V(0)$ observed in the QAH phase ($c=\pm1$) [Fig. \ref{fig2}(e)], which stands in stark contrast to the linear distribution found in the axion insulator ($c=0$) regime.
This exponential voltage profile can be experimentally measured using a multi-terminal Hall bar device depicted in Fig.~\ref{fig2}(f).
Furthermore, local heat dissipation $P_D$ [Figs.~\ref{fig2}(g) and (h)] reveals an exponential localization at the sample edges for the QAH state, in sharp contrast to the uniform distribution observed in the axion insulator state.
A detailed quantitative comparison between theory and numerical results, presented in the Supplemental Material \cite{SM}, shows excellent agreement across different system parameters.

The good agreement  between our Landauer-B{\"u}ttiker simulations (in Fig.~\ref{fig2}) and analytic theory (in Fig.~\ref{fig1})  underscores the reliability of our findings and the experimental feasibility of observing the NHSE in magnetic TIs. Recent experiments have already demonstrated magnetic switching between QAH and axion insulator phases via external fields in both magnetically doped (Bi,Sb)$_2$Te$_3$ TI sandwich structures and MnBi$_2$Te$_4$ \cite{Mogi2017,XiaoDi2018,liu2020robust}.
Moreover, the interplay between chiral and normal edge states has been shown to be a primary dissipation mechanism in these magnetic TI heterostructures \cite{Chang2015,Charles2020,Chang2023} (see, e.g., Fig.~4o in the experimental work of Chang \textit{et al.}~\cite{Chang2015}) . This suggests that the proposed intrinsic Non-Hermitian topology is well within the reach of current experimental techniques.

\par
\vspace{1em}

\noindent\textbf{\large Discussion}\\
We emphasize that the exponential behavior discussed here is fundamentally different from Anderson localization. The latter is characterized by the Anderson localization length $\xi$, which describes exponential decay of single-particle wavefunctions due to phase-coherent quantum interference. In contrast, our mechanism originates from nonreciprocal transport of the edge modes governed by the continuum Hatano--Nelson equation, introducing a characteristic length $\lambda$ that governs the spatial evolution of the electrochemical potential.
The proposed non-Hermitian mechanism can be experimentally distinguished from Anderson localization by two criteria. First, exponential decay of conductance in our model appears only in the QAH state but not in the axion-insulator state of the same device in the classical transport regime. This exponential behavior persists while the Anderson localization is suppressed by dephasing when increasing temperature. Second, reversing magnetization reverses edge chirality and thus the direction of exponential voltage and dissipation profiles, providing a chirality-dependent signature absent in conventional Anderson localization.
Notably, in the ideal QAH limit with only a chiral edge mode ($\sigma_D=0$), the exponential voltage and heat-dissipation profiles disappear.

In summary, we have shown that a Hermitian quantum anomalous Hall system can intrinsically host non-Hermitian topology through the interplay between chiral and diffusive edge modes. 
Without introducing externally imposed dissipation or contact engineering, the edge transport dynamics naturally reduce to a continuum Hatano-Nelson model, leading to an experimentally measurable non-Hermitian skin effect manifested in voltage distribution, heat dissipation, and exponential longitudinal conductance.
Our findings reveal that non-Hermitian topology need not rely on  externally engineered gain/loss or reservoir-induced non-Hermitian band structures, but can instead emerge from microscopic chiral transport processes within a globally Hermitian quantum platform.
This intrinsic mechanism provides a realistic and scalable route toward observing non-Hermitian phenomena in solid-state devices. 
Given the rapid experimental progress in quantum anomalous Hall heterostructures, the predicted exponential transport signatures should be directly accessible with existing measurement techniques.
More broadly, this work establishes a bridge between mesoscopic transport theory and non-Hermitian topology, and opens new avenues for designing topological nonreciprocal electronic functionalities based on quantum materials.
\\
\\
\noindent \textbf{\large{Methods}}\\
\textbf{Numerical Method for the Voltage and Heat Dissipation in Magnetic Topological Insulators}\\
\begin{figure}[bht]
	\centering
	\includegraphics[width=3.2in]{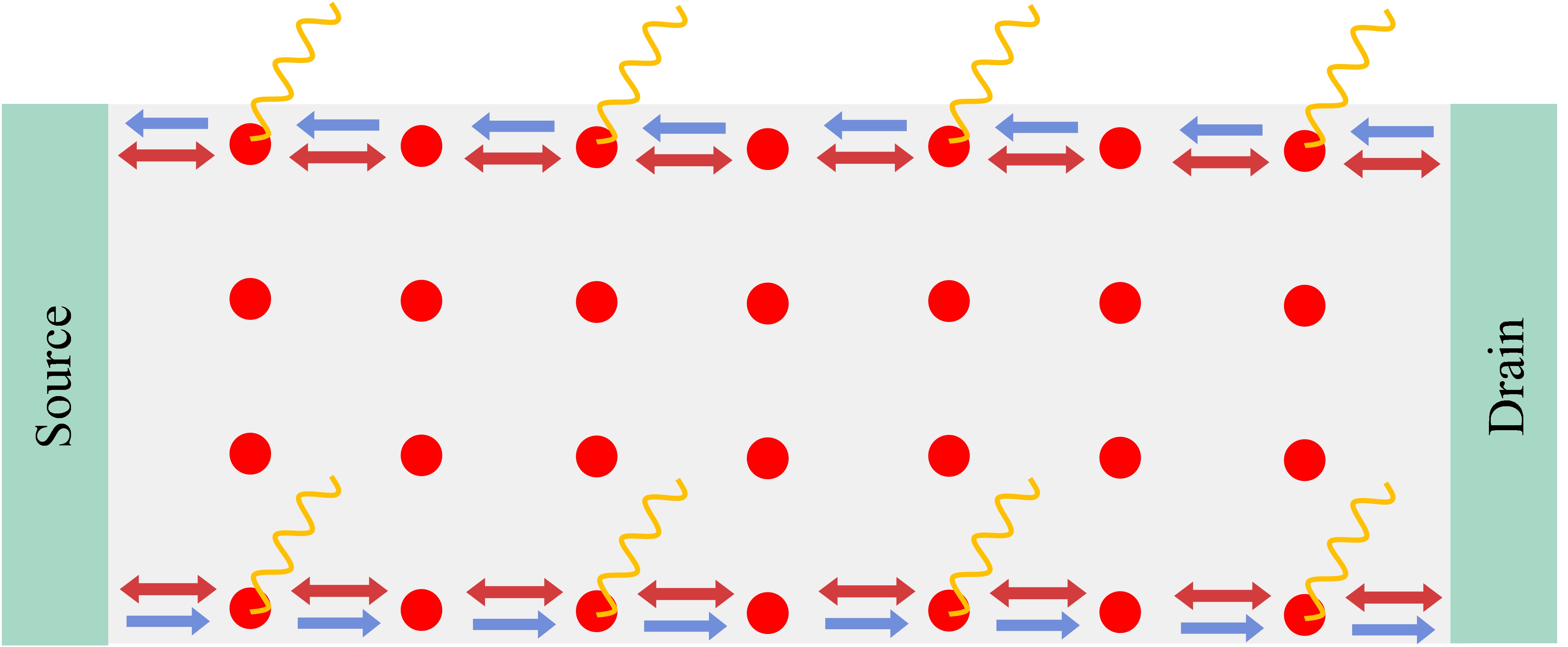}
	\caption{ 
		Schematic illustration of the lattice model obtained from discretizing the QAH Hamiltonian.
		Unidirectional blue arrows indicate the chiral channel originating from the chiral edge state,
		while bidirectional red arrows denote the normal channel arising from the normal edge state.
		The curved lines represent virtual leads attached to the corresponding lattice sites.
		\label{figs2}}
\end{figure}
We discretize the Hamiltonian $H_{\text{MTI}}$ in the main text on cubic lattices:
\begin{eqnarray}\label{eq11}
	H_{\text{MTI}}=&\left[
	\sum_{{\bf i}}{\bf c}_{\bf i}^{\dagger}{\mathcal M}_0{\bf c}_{\bf i}
	+\left(\sum_{{\bf i},j=x,y,z}{\bf c}_{\bf i}^{\dagger}{\mathcal T}_j{\bf c}_{{\bf i}+{\bf \hat{e}}_j}+\text{H.c.}\right)
	\right]
	+\sum_{\bf i} {\bf c}_{\bf i}^{\dagger}{\mathcal M}_z{\bf c}_{\bf i} \nonumber
\end{eqnarray}
where ${\mathcal M}_0=\left(M_0-\sum_{j=x,y,z}2B_j/a^2\right)\sigma_z{\otimes}s_0$, ${\mathcal M}_z=M_z(z)\sigma_0{\otimes}s_z$, and ${\mathcal T}_j=B_j/a^2\sigma_z{\otimes}s_0-iA_j/(2a)\sigma_x{\otimes}s_j$. Here, $a=1$ is the cubic lattice constant.
${\bf c}_{\bf i}$ (${\bf c}_{\bf i}^{\dagger}$) is the annihilation (creation) operator at site ${\bf i}$, and ${\bf \hat{e}}_j$ is a unit vector in the $j$ direction for $j=x,y,z$.
The sample width, length, and thickness are given by 
$W = N_y a$, $L = N_x a$, and $L_z = N_z a$, respectively.

We study the edge transport behavior in a six-terminal Hall-bar device. 
Within the Landauer--B{\"u}ttiker formalism\cite{Buttiker1988,Data1995}, 
the electric current and the heat current are given by 
\begin{align}
	&I_p = \frac{e^2}{h} \sum_{q \neq p} \left( T_{qp}V_p - T_{pq}V_q \right),\nonumber
	\\
	&P_D(p) = \frac{e^2}{2h} \sum_{q \neq p} T_{qp} \left(V_p - V_q \right)^2,
\end{align}
where $V_p$ is the voltage at lead $p$. $I_p$ denotes the current flowing out of lead $p$, while $P_D(p)$ denotes the heat dissipation at lead $p$. The transmission probability from lead $q$ to lead $p$ is expressed as $T_{pq}(E_F) = \mathrm{Tr}\left[ {\bm \Gamma}_p {\bm G}^r {\bm \Gamma}_q {\bm G}^a \right]$, with the linewidth function ${\bm \Gamma}_p = i({\bm \Sigma}_p^r - {\bm \Sigma}_p^{r\dagger})$ and the retarded Green's function ${\bm G}^r = [E_F{\bm I} - H_{\text{MTI}} - \sum_p {\bm \Sigma}_p^r]^{-1}$. Here ${\bm \Sigma}_p^r$ is the self-energy due to the coupling between the central region and lead $p$.
For system sizes exceeding the dephasing length, the transport behavior is well consistent with semiclassical Boltzmann transport theory.
The dephasing effect and heat dissipation is modeled by attaching B{\"u}ttiker virtual leads to the side surfaces~\cite{Fang2021,Yanxia2008,Buttiker1986,Buttiker1988} (see Fig.~\ref{figs2}).
 Note that we use the wide-band approximation for the real and virtual leads with ${\bm \Gamma}_p=\Gamma_p {\bf I}_p$, where $\Gamma_p$  is the dephasing strength\cite{Yanxia2008}. ${\bf I}_p$ is $4n_p \times 4n_p$ unit matrix. In our model, $\Gamma_{p}=2.0$ for source and drain, and $\Gamma_p = 0.4$ for other leads.
$n_p$ is the number of sites coupling to the lead $p$, where $n_{p}=N_yN_z$ for the source and drain leads and $n_p=N_z$ for other leads. 

In our calculations, the source and drain leads act as current electrodes with fixed voltage $\pm V_0$ and other leads act as voltage electrodes. Combining Landauer-B{\"u}ttiker formula with those boundary conditions in the leads, one can calculate the voltage of each lead and the longitudinal current $I$. 
$V(x)$ denotes the voltage of the virtual lead located at position $x$ along the one-dimensional coordinate.
The longitudinal and Hall resistances are obtained from the measured voltages as $\rho_{xx} = [V(x)-V(x+L_0)]W/(IL_0)$ and $\rho_{xy} = [V(x)-V(-x)]/I$. The corresponding longitudinal and Hall conductance are obtained from the tensor relation: $\sigma_{xx}=\rho_{xx}/(\rho_{xx}^2+\rho_{xy}^2)$, and $\sigma_{xy}=\rho_{xy}/(\rho_{xx}^2+\rho_{xy}^2)$. 
Only electrons in the edge states participate in the electrical transport when $E_F$ is located in the bulk gap at zero or low temperature. Therefore, we just need to consider the dephasing process on the edge. We simulate the dephasing process by using $2n_x$ uniformly distributed B{\"u}ttiker's virtual leads on top and bottom, when we calculate the voltage and conductance.
The heat dissipation distribution is calculated by using $n_x\times n_y$ uniformly distributed B{\"u}ttiker's virtual leads on the sample.
Throughout our calculations, we set $n_i/N_i = 0.2$.
$N_i$ and $n_i$ are the number of total sites and selected sites in the $i$ direction for $i=x,y$, respectively.
\\
\\
\noindent \textbf{Data availability}\\
The source data underlying the main and supplementary figures are provided with this paper as a Source Data file. Additional data generated in this study are available from the corresponding authors upon reasonable request. 

\par
\vspace{1em}

\noindent \textbf{Code availability}\\
The computation code for getting the theoretical predictions is available from the corresponding authors upon reasonable request.

\bibliographystyle{naturemag}
\bibliography{ref1}

\noindent \textbf{Acknowledgements}\\
We thank Qing-Feng Sun, Hailong Li, Qing Yan, Hua Jiang for helpful and illuminating discussions.
This work was financially supported by  the National Key R and D Program of China (Grant No. 2025YFA1412100), the National Natural Science Foundation of China (Grants Nos. 123B2058, Nos.12574043), and the Natural Science Foundation of Jiangsu Province Grant (BK20230066). X.C.X. acknowledges support from the Innovation Program for Quantum Science and Technology (grant no. 2021ZD0302400)
\\
\\
\noindent \textbf{Author contributions}\\	
X.-C. X. and C.-Z. C. conceived the idea and supervised the research. H. Z. performed the analytical and numerical calculations with the help of X.-C. X., C.-Z. C. and M. L.. H. Z. and C.-Z. C wrote the manuscript. All authors discussed the results and commented on the manuscript.
\\
\\
\noindent \textbf{Competing interests}\\
The authors declare no competing interests.
	
\end{document}


\title{Supplementary Materials for ``Non-Hermitian Topology from Edge Transport in Hermitian Quantum Anomalous Hall Systems''}
\author{Humian Zhou}
\affiliation{International Center for Quantum Materials, School of Physics, Peking University, Beijing 100871, China}
\author{Ming Lu}
\affiliation{Beijing Academy of Quantum Information Sciences, Beijing 100193, China}
\author{Chui-Zhen Chen}
\email{czchen@suda.edu.cn}
\affiliation{School of Physical Science and Technology, Soochow University, Suzhou 215006, China}
\affiliation{Institute for Advanced Study, Soochow University, Suzhou 215006, China}
\author{X. C. Xie}
\email{xcxie@pku.edu.cn}
\affiliation{International Center for Quantum Materials, School of Physics, Peking University, Beijing 100871, China}
\affiliation{Institute for Nanoelectronic Devices and Quantum Computing, Fudan University, Shanghai 200433, China}
\affiliation{Hefei National Laboratory, Hefei 230088, China}
\date{\today }

\maketitle
\tableofcontents
	\section{Quantum Kinetic Theory for the Edge transport}
	\subsection{Quantum Kinetic Equation}
	We consider a multiband system described by the Hamiltonian 
	\begin{align}
		\hat{H}=\hat{H}_0+U(\hat{r})+U_D(\hat{r}).
	\end{align}
Here, $\hat{H}_0$ denotes the effective n-band Hamiltonian, and $U(\hat{r})$ represents a slowly varying electrostatic potential. $U_D(\hat{r})$ is the disorder potential, and is given by 
	\begin{align}
		U_D(\hat{r})=\sum_{i,\alpha}U_{\alpha }(\hat{r}-R_{i}),
	\end{align}
where $R_i$ denote the positions of randomly distributed impurities($i=1,\dots,N$ where $N$ the total number of impurities in the volume $V$), and $U_{\alpha}(r)$ represents the $\alpha$-type scattering potential of a single impurity placed at the origin of the coordinate system.
	The density operator $\hat{\rho}_t$ satisfies the quantum Liouville equation
	\begin{align}
		\frac{\partial \hat{\rho}_t}{\partial t}+\frac{i}{\hbar}[\hat{H},\hat{\rho}_t]=0.
	\end{align}
After disorder averaging, we obtain the equation 
	\begin{align}\label{eq_r}
		\frac{\partial \langle \hat{\rho}_t  \rangle}{\partial t}
		+\frac{i}{\hbar}[\hat{H}_0+U(\hat{r}),\langle \hat{\rho}_t \rangle]
		+ \frac{i}{\hbar}\sum_{i,\alpha} 
		\int \frac{\mathrm{d}R_{i}}{V}
		[U_{\alpha}(\hat{r}-R_{i}), \langle \hat{\rho}_t \rangle_{i} ]=0,
	\end{align}
	and
	\begin{align}
		\frac{\partial \langle \hat{\rho}_t  \rangle_{i}}{\partial t}
		+\frac{i}{\hbar}[\hat{H}_0+U(\hat{r})+U_{\alpha}(\hat{r}-R_{i}),\langle \hat{\rho}_t \rangle_{i}]
		+\frac{i}{\hbar}\sum_{j\neq i,\alpha}
		\int \frac{\mathrm{d}R_{j}}{V}
		[U_{\alpha}(\hat{r}-R_{j}), \langle \hat{\rho}_t \rangle_{{ij}} ]=0.
	\end{align}	
Here,
\begin{align}
		\langle \hat{\rho}_t  \rangle &= \int \frac{\mathrm{d}R_1}{V} \cdots \int \frac{\mathrm{d}R_N}{V}\hat{\rho}_t, \\
		\langle \hat{\rho}_t  \rangle_{i} &= \int \frac{\mathrm{d}R_1}{V} \cdots
		\int \frac{\mathrm{d}R_{i-1}}{V} 
		\int \frac{\mathrm{d}R_{i+1}}{V}
		\cdots 
		\int \frac{\mathrm{d}R_N}{V}\hat{\rho}_t, \\
		\langle \hat{\rho}_t  \rangle_{ij} &= \int \frac{\mathrm{d}R_1}{V} \cdots
		\int \frac{\mathrm{d}R_{i-1}}{V} 
		\int \frac{\mathrm{d}R_{i+1}}{V}
		\cdots 
		\int \frac{\mathrm{d}R_{j-1}}{V} 
		\int \frac{\mathrm{d}R_{j+1}}{V}
		\cdots
		\int \frac{\mathrm{d}R_N}{V}\hat{\rho}_t.
\end{align}
We define the correlation operator $\hat{\kappa}_{i}= \langle \hat{\rho}_t \rangle_{i} - \langle \hat{\rho}_t \rangle$. 
In the first order in the electron-impurity interaction, this correlation operator satisfies\cite{Vasko2005}
\begin{align}\label{eq_k}
		\frac{\partial \langle \kappa \rangle_{i}}{\partial t}
		+\frac{i}{\hbar}[\hat{H}_0+U(\hat{r}),\hat{\kappa}_{i}]
		+\frac{i}{\hbar}[U_{\alpha}(\hat{r}-R_{i}),\langle \hat{\rho}_t \rangle]=0.
\end{align}
The solution of this equation is
\begin{align}\label{eq_k}
		\kappa_{i} 
		= \frac{1}{i \hbar} \int_{-\infty}^{t} dt' e^{\eta t'} 
		\hat{S}(t, t') [U_{\alpha}(\hat{r}-R_{i}), \langle\hat{\rho}_{t'}\rangle] 
		\hat{S}^{\dagger}(t, t'),
\end{align}
where $\eta \rightarrow 0^+$, and the evolution operator is 
\begin{align}
	\hat{S}(t, t') = e^{-\frac{i}{\hbar}(\hat{H}_0+U(\hat{r}))(t-t')}.
\end{align}
Substituting Eq.~\ref{eq_k} back into Eq.~\ref{eq_r}, we obtain the quantum kinetic equation
\begin{align}\label{eq:qL}
	\frac{\partial \langle \hat{\rho}_t \rangle}{\partial t}
	+\frac{i}{\hbar}[\hat{H}_0+U(\hat{r}),\langle \hat{\rho}_t \rangle]
	= J(\hat{\rho}),
\end{align}
where the scattering term reads
\begin{align}
	J(\hat{\rho})=\frac{1}{\hbar^2}\sum_{i} 
	\int \frac{\mathrm{d}R_{i}}{V}\int_{-\infty}^{t} dt' e^{\eta t'} 
	[\hat{S}(t, t') [U_{\alpha}(\hat{r}-R_{i}), \langle\hat{\rho}_{t'}\rangle] 
	\hat{S}^{\dagger}(t, t'),U_{\alpha}(\hat{r}-R_{i})].
\end{align}

\subsection{Wigner Transformation and Boltzmann Equation}

Applying the Wigner transformation,
\begin{align}
	\rho^{nm}_k(r,t)=\int \mathrm{d}R\, e^{-ikR}
	\bra{n,r+R/2} \langle \hat{\rho}_t \rangle \ket{m,r-R/2},
\end{align}
the quantum kinetic equation in Eq.~(\ref{eq:qL}) can be recast as\cite{Sekine2017,Culcer2017}
\begin{align} \label{eq:qke}
	\frac{\partial \rho_k}{\partial t} 
	+\frac{i}{\hbar}[H_0,\rho_k]
	+\frac{1}{2}\left\{\frac{\mathrm{D}H_0}{\hbar\mathrm{D}k},\nabla \rho_k\right\}
	-\nabla U \frac{\mathrm{D}\rho_k}{\hbar\mathrm{D}k}=J_k,
\end{align}
where the gauge-covariant derivative is defined as
\begin{align}
	\frac{\mathrm{D}A}{\mathrm{D}k}
	=\nabla_k A - i[\mathcal{R},A],
\end{align}
and $\mathcal{R}$ denotes the Berry connection. 

The scattering term reads
\begin{align}
	J_k^{nn'} &= \frac{i}{\hbar} \sum_{\alpha} n_{\alpha} 
	\sum_{n'', n'''} \int \frac{\mathrm{d}k'}{2\pi} \Bigg[
	\frac{
		U^{nn''}_{\alpha kk'} U^{n''n'''}_{\alpha k'k}\rho^{n'''n'}_{k'} 
		- U^{nn''}_{\alpha kk'} \rho^{n''n'''}_{k'} U^{n'''n'}_{\alpha k'k} 
	}{
		\varepsilon^{n''}_{k'} - \varepsilon^{n'}_k - i\hbar \eta
	}  \notag \\[3pt]
	&\quad + 
	\frac{
		\rho^{nn'''}_{k} U^{n'''n''}_{\alpha kk'} U^{n''n'}_{\alpha k'k} 
		- U^{nn'''}_{\alpha kk'} \rho^{n'''n''}_{k'} U^{n''n'}_{\alpha k'k}
	}{
		\varepsilon^{n}_k - \varepsilon^{n''}_{k'} - i\hbar \eta
	} 
	\Bigg],
\end{align}
where $n_{\alpha}$ is the impurity density, and 
$U^{nn''}_{\alpha kk'} = \bra{nk}U_{\alpha}(\hat{r})\ket{n''k'}$ denotes the matrix element of the impurity potential.   $\ket{nk}$ and $\varepsilon_k^n$ are the eigenvector and eigenenergy of $H_0$,respectively.

To analyze the contributions from different components of the density matrix, 
we decompose $\rho_k^{nn'}$ into diagonal and off-diagonal parts,
\begin{align}
	\rho_k^{nn'} = \delta_{nn'} f_k^n + (1 - \delta_{nn'}) S_k^{nn'},
\end{align}
and correspondingly separate the scattering term as
\begin{align}
	J_k^{nn'} = J_{k,f}^{nn'} + J_{k,S}^{nn'},
\end{align}
with $J_{k,f}^{nn'}$ and  $J_{k,S}^{nn'}$ containing the contributions of the $f_k^n$ and $S_k^{nn'}$, respectively.
The diagonal component yields
\begin{align}
	J_{k,f}^{nn}
	= - \sum_{n'} \int \frac{\mathrm{d}k'}{2\pi}\,
	W_{nk,n'k'} \big( f^{n}_{k} - f^{n'}_{k'} \big),
\end{align}
where the scattering rate is given by
\begin{align} \label{eq_W}
	W_{nk,n'k'}
	= \frac{2\pi}{\hbar} \sum_{\alpha} n_{\alpha}\,
	U^{nn'}_{\alpha kk'} U^{n'n}_{\alpha k'k}\,
	\delta\!\left( \varepsilon^{n}_k - \varepsilon^{n'}_{k'} \right).
\end{align}

For notational convenience, we henceforth place the band index in the subscript rather than the superscript. The Boltzmann equation for the diagonal component of Eq.~(\ref{eq:qke}) reads
\begin{align}
	\frac{\partial f_{nk}}{\partial t} 
	+ v_{nk} \nabla f_{nk}
	- \nabla U \frac{\partial f_{nk}}{\hbar \partial k}
	= J_{k,f}^{nn} + J_{k,S}^{nn},
\end{align}
where $v_{nk} = \frac{\partial \varepsilon_{nk}}{\hbar \partial k}$ denotes the group velocity. 
We consider a weakly nonequilibrium state and write the distribution function as
$f_{nk} \simeq f_{nk}^{(0)} + f_{nk}^{(1)}$,
where $f_{nk}^{(0)} = f_0(\varepsilon_{nk} - \mu)$ is the equilibrium Fermi--Dirac distribution and $\mu(x)=\tilde{\mu}(x)-U(x)$ is the local chemical potential.
The off-diagonal scattering term $J_{k,S}^{nn'}$ does not contribute to the first-order correction $f_{nk}^{(1)}$ \cite{Ma2025}. Therefore, we get 
\begin{align}\label{eq_scattering}
	\frac{\partial f_{nk}}{\partial t} 
	+ v_{nk} \nabla f_{nk}
	- \nabla U \frac{\partial f_{nk}}{\hbar \partial k} 
= - \sum_{n'} \int \frac{\mathrm{d}k'}{2\pi  }\,
	W_{nk,n'k'} \big( f_{nk} - f_{n'k'} \big)
\end{align}

\subsection{First-Order Correction to the Distribution Function}

we consider the system in equilibrium, where the distribution function $f_{nk}$ takes the Fermi--Dirac form 
$f_0(\varepsilon_{nk} - \mu_0)$. 
Substituting this into the Boltzmann equation gives
\begin{align}
	\frac{\partial f_0(\varepsilon_{nk}-\mu_0)}{\partial t} 
	+ v_{nk} \nabla f_0(\varepsilon_{nk}-\mu_0)
	- \nabla U \frac{\partial f_0(\varepsilon_{nk}-\mu_0)}{\hbar\,\partial k}
	= 0.
\end{align}
The above condition requires $\nabla(\mu_0 + U) = 0$
which implies that the  electrochemical potential $\tilde{\mu}_0 = \mu_0 + U$ remains spatially uniform at equilibrium. Here, $\mu_0$ is the chemical potential at equilibrium.

For the non-equilibrium steady state,  
$f_{nk} = f_{nk}^{(0)} + f_{nk}^{(1)}$ with $f_{nk}^{(0)} = f_0(\varepsilon_{nk} - \mu)$.
The Boltzmann equation then reads
\begin{align}
	v_{nk} \nabla f_{nk} 
	- \nabla U \frac{\partial f_{nk}}{\hbar \partial k}  
	= - \sum_{n'} \int \frac{\mathrm{d}k'}{2\pi} \, W_{nk,n'k'} \left( f_{nk} - f_{n',k'} \right).
\end{align}
or 
\begin{align}
	v_{nk} \nabla f_{nk}^{(0)} + v_{nk} \nabla f_{nk}^{(1)} 
	- \nabla U \frac{\partial f_{nk}^{(0)}}{\hbar \partial k} 
	- \nabla U \frac{\partial f_{nk}^{(1)}}{\hbar \partial k} 
	= - \sum_{n'} \int \frac{\mathrm{d}k'}{2\pi} \, W_{nk,n'k'} \left( f_{nk}^{(1)} - f_{n'k'}^{(1)} \right).
\end{align}
Equivalently, introducing the matrix $M_{n k, n' k'} = \sum_{n''} \int {\mathrm{d}k''} W_{n k, n'' k''} \delta_{n n'} \delta(k - k') - W_{n k, n' k'}$, 
\begin{align} \label{eq_27}
	- v_{nk} \frac{\partial f_{nk}^{(0)}}{\partial \varepsilon} \nabla \tilde{\mu}
	+ v_{nk} \nabla f_{nk}^{(1)} 
	- \nabla U \frac{\partial f_{nk}^{(1)}}{\hbar \partial k} 
	= -\sum_{n'} \int \frac{\mathrm{d}k'}{2\pi} \, M_{nk,n'k'} f^{(1)}_{n'k'}.
\end{align}
where $\tilde{\mu}(x)=\mu(x) + U(x)$ is the electrochemical potential.
For elastic collisions, the first-order correction $f_{nk}^{(1)}$ satisfies
\begin{align} \label{eq_28}
	\sum_{n,n'} \int \frac{\mathrm{d}k}{2\pi}\,\frac{\mathrm{d}k'}{2\pi} \, \delta(\varepsilon-\varepsilon_{nk}) \, M_{nk,n'k'} f^{(1)}_{n'k'} = 0.
\end{align}
Consequently, taking the sum over $n$ and the integral over $k$ on both sides of Eq.~\ref{eq_27}, we obtain the energy-resolved current balance that reads
\begin{align}
	\sum_{n} \int \frac{\mathrm{d}k}{2\pi} \, \delta(\varepsilon-\varepsilon_{nk}) 
	\Big( v_{nk} \nabla f_{nk}^{(1)} - \nabla U \frac{\partial f_{nk}^{(1)}}{\hbar \partial k} \Big)
	= \sum_{n} \int \frac{\mathrm{d}k}{2\pi} \, \delta(\varepsilon-\varepsilon_{nk}) \, v_{nk} 
	\frac{\partial f_{nk}^{(0)}}{\partial \varepsilon} \nabla \tilde{\mu}
	= g(\varepsilon)\, \frac{\partial f_0(\varepsilon-\mu)}{\partial \varepsilon} \, \bar{v}(\varepsilon) \nabla \tilde{\mu}.
\end{align}
Here, $g(\varepsilon) = \sum_n g_n(\varepsilon)$ is the total density of states, where $g_n(\varepsilon) = \int \frac{\mathrm{d}k}{2\pi} \, \delta(\varepsilon - \varepsilon_{nk})$
is the density of states of the $n$th band.
The average group velocity at energy $\varepsilon$ is defined as
\begin{align}
	\bar{v}(\varepsilon)
	= \frac{1}{g(\varepsilon)} \sum_n \int \frac{\mathrm{d}k}{2\pi} \,
	\delta(\varepsilon - \varepsilon_{nk}) \, v_{nk}.
\end{align}
Then, the first-order correction $f_{nk}^{(1)}$ satisfies
\begin{align}
	& -\frac{\partial f_{nk}^{(0)}}{\partial \varepsilon} \left( v_{nk} - \bar{v}(\varepsilon_{nk}) \right) \nabla \tilde{\mu}
	+ v_{nk} \nabla f_{nk}^{(1)} 
	- \nabla U \frac{\partial f_{nk}^{(1)}}{\hbar \, \partial k} \notag \\
	& \quad - g(\varepsilon_{nk})^{-1}\sum_{n''} \int \frac{\mathrm{d}k'}{2\pi} \, \delta(\varepsilon_{nk} - \varepsilon_{n''k'}) 
	\left( v_{n''k'} \nabla f^{(1)}_{n''k'} - \nabla U \frac{\partial f^{(1)}_{n''k'}}{\hbar \, \partial k} \right)
	= -\sum_{n'} \int \frac{\mathrm{d}k'}{2\pi} \, M_{nk,n'k'} f^{(1)}_{n'k'}.
\end{align}
Therefore, the first-order correction to the distribution function reads
\begin{align}
	f_{nk}^{(1)} 
	&=  \sum_{n'} \int \frac{\mathrm{d}k'}{2\pi} \, [M^{-1}]_{nk,n'k'} 
	\Big( v_{n'k'} - \bar{v}(\varepsilon_{n'k'}) \Big) 
	\frac{\partial f_0(\varepsilon_{n'k'}-\mu)}{\partial \varepsilon} \, \nabla \tilde{\mu} \notag \\
	& \quad - \sum_{n'} \int \frac{\mathrm{d}k'}{2\pi} \, [M^{-1}]_{nk,n'k'}
	\Big[ v_{n''k} \nabla f^{(1)}_{n''k} - g(\varepsilon_{nk})^{-1} \sum_{n''} \int \mathrm{d}k \, \delta(\varepsilon-\varepsilon_{n''k}) v_{n''k} \nabla f^{(1)}_{n''k} \Big] \notag \\
	& \quad + \nabla U \sum_{n'} \int \frac{\mathrm{d}k'}{2\pi} \, [M^{-1}]_{nk,n'k'} 
	\Big[ \frac{\partial f_{nk}^{(1)}}{\hbar \, \partial k} - g(\varepsilon_{nk})^{-1}\sum_{n''} \int \mathrm{d}k \, \delta(\varepsilon-\varepsilon_{n''k}) \frac{\partial f^{(1)}_{n''k}}{\hbar \, \partial k} \Big].
\end{align}
Since the electrochemical potential varies slowly in space, we retain only the first-order terms in $\nabla \tilde{\mu}$ and neglect higher-order contributions such as $\nabla^2 \tilde{\mu}$ or $(\nabla \tilde{\mu})^2$. Consequently, the expression simplifies to
\begin{align} \label{eq_f1}
	f_{nk}^{(1)} 
	=  \sum_{n'} \int \frac{\mathrm{d}k'}{2\pi} \, [M^{-1}]_{nk,n'k'} 
	\Big( v_{n'k'} - \bar{v}(\varepsilon_{n'k'}) \Big) 
	\frac{\partial f_0(\varepsilon_{n'k'}-\mu)}{\partial \varepsilon} \, \nabla \tilde{\mu}.
\end{align}
Note that the total density response, obtained by summing over all bands and momenta, corresponds to a zero eigenvalue of the collision matrix $M_{nk,n'k'}$. Consequently, the matrix $M$ is not strictly invertible, since its inverse $M^{-1}_{nk,n'k'}$ would be ill-defined along this zero mode. In practice, this issue is resolved by projecting out the subspace corresponding to the total density, i.e., by decomposing $M$ and $M^{-1}$ so that the zero mode is removed from the relevant vector space\cite{Culcer2017}.

\subsection{Electric Current, Energy Current, and Heat Dissipation}
The electric current can be expressed as
\begin{align}
	j &= -e\sum_n \int \frac{\mathrm{d}k}{2\pi}\, v_{nk} f_{nk} \notag\\
	&= -e\sum_n \int \frac{\mathrm{d}k}{2\pi}\, \bar{v}(\varepsilon_{nk}) f_0(\varepsilon_{nk}-\mu)
	-e\sum_n \int \frac{\mathrm{d}k}{2\pi}\, (v_{nk}-\bar{v}(\varepsilon_{nk})) f_{nk}^{(1)}  \notag\\
	&= -e\sum_{m} \int \frac{\mathrm{d}k}{2\pi}\, v_{nk} [f_0(\varepsilon_{nk}-\mu)-f_0(\varepsilon_{nk}-\mu_0)] + j_0 + \frac{\sigma_D}{e} \nabla \tilde{\mu},
\end{align}
where $j_0 = -e\sum_{n}\int \frac{\mathrm{d}k}{2\pi}\, v_{nk} f_0(\varepsilon_{nk}-\mu_0)$ denotes the equilibrium current, which does not contribute to charge transport.

The longitudinal conductance is then given by
\begin{align}
	\sigma_D
	=  e^2\sum_{n,n'} \int \frac{\mathrm{d}k}{2\pi}\,\frac{\mathrm{d}k'}{2\pi}\,
	(v_{nk}-\bar{v}(\varepsilon_{nk})) [M^{-1}]_{nk,n'k'} (v_{n'k'}-\bar{v}(\varepsilon_{n'k'}))
	\left(-\frac{\partial f_0(\varepsilon_{n'k'}-\mu)}{\partial\varepsilon_{n'k'}}\right).
\end{align}
Finally, the continuity equation takes the form
\begin{align}\label{eq_ce}
	\frac{\partial \rho}{\partial t} + \nabla \cdot (j - j_0) = 0,
\end{align}
which ensures the conservation of charge after removing the equilibrium current contribution. The charge density is given by $\rho = -e\sum_n \int \frac{\mathrm{d}k}{2\pi}\, f_{nk}$. In the present of a source term $S$, Eq.~\ref{eq_ce} becomes
\begin{align}\label{eq_ce2}
	\frac{\partial \rho}{\partial t} + \nabla \cdot (j - j_0) = S,
\end{align}

The energy current is given by:
\begin{align}\label{eq_ec}
	j_E &= \sum_n \int \frac{\mathrm{d}k}{2\pi}\, v_{nk} (\varepsilon_{nk}+U) f_{nk} \notag\\
\end{align}
and the heat dissipation is given by:
\begin{align}
	P_D &= \nabla \cdot (j_E-j_{0,E}) \notag\\
\end{align}
where $j_{0,E}$ is the energy current at equilibrium.

\subsection{Two-Band Model}
We consider a two-band Hamiltonian
\begin{align}
	H_0=\begin{pmatrix}\varepsilon_{1k} & 0  \\ 0 & \varepsilon_{2k} \end{pmatrix},
\end{align}
with dispersions $\varepsilon_{1,k} =\varepsilon_{1,-k}$ (normal mode) and $\varepsilon_{2k} = c\hbar v_c k$ (chiral mode). The parameter $v_c$ denotes the chiral mode velocity, and  $c=\pm1$ indicates the propagation direction.
The single-particle scattering potential is
\begin{align}
	U_{\alpha}(\hat{r})=u_{\alpha}\sigma_{\alpha} \delta(\hat{r}),
\end{align}
where the Pauli matrices \( \sigma_{\alpha} \) ( $\alpha=0,1$) act on the band indices. The parameters $u_{0}$ and $u_{1}$ represent the intra-band and inter-band scattering strengths, respectively. 
Then, according to Eq.~\ref{eq_W}, the scattering rate is given by
\begin{equation}\label{eq_W2}
	W_{nk,n'k'} = \frac{2\pi}{\hbar} \sum_{\alpha} n_\alpha u_\alpha^2 \,
	(\sigma_{\alpha})_{nn'} (\sigma_{\alpha})_{n'n} \,
	\delta\bigl(\varepsilon_{nk} - \varepsilon_{n'k'}\bigr)
\end{equation}
According to Eq.~\ref{eq_f1}, the first-order corrections to $f_{nk}$ are:
\begin{equation}\label{eq_f}
	\begin{cases}
		f^{(1)}_{1k} = \left[ v_{1k} \tau_1(\varepsilon_{1k}) - v_{2k} \tau_2(\varepsilon_{1k}) \dfrac{g_2(\varepsilon)}{g_1(\varepsilon)} \right] \partial_{\varepsilon} f_0(\varepsilon_{1k} - \mu) \, \nabla \tilde{\mu}, \\[12pt]
		f^{(1)}_{2k} = v_{2k} \tau_2(\varepsilon_{2k}) \, \partial_{\varepsilon} f_0(\varepsilon_{2k} - \mu) \, \nabla \tilde{\mu}.
	\end{cases}
\end{equation}
Here, $\tau_1$ and $\tau_2$ is the relaxation time for the distributions of normal edge channel ($f_{1k}$) and the chiral edge channel ($f_{2k}$) to reach equilibrium.
The explicit forms of the relaxation time are 
\begin{equation}\label{eq_tau}
	\begin{cases}
		\dfrac{\hbar}{\tau_1(\varepsilon)} = 2\pi \left[ n_0 u_0^2 g_1(\varepsilon) + n_1 u_x^2 g_2(\varepsilon) \right], \\[6pt]
		\dfrac{\hbar}{\tau_2(\varepsilon)} = 2\pi n_1 u_x^2 \dfrac{\bigl[ g_1(\varepsilon) + g_2(\varepsilon) \bigr]^2}{g_1(\varepsilon)},
	\end{cases}
\end{equation}
where $g_n(\varepsilon) = \int \mathrm{d}k \, \delta(\varepsilon - \varepsilon_{nk})$ is the density of states of band $n$. At low temperature, the derivative of the Fermi-Dirac function is sharply peaked at the Fermi energy and hence $f_{nk}^{(1)}$ can be written as:
\begin{equation}\label{eq_f}
	\begin{cases}
		f^{(1)}_{1k} = \left[ \operatorname{sign}(k) v_{1F} \tau_1(E_F) - c v_{2F} \tau_2(E_F) g_2/g_1\right] \partial_{\varepsilon} f_0(\varepsilon_{1k} - E_F) \, \nabla \tilde{\mu}, \\[12pt]
		f^{(1)}_{2k} = c v_{2F} \tau_2(E_F) \, \partial_{\varepsilon} f_0(\varepsilon_{2k} - E_F) \, \nabla \tilde{\mu}.
	\end{cases}
\end{equation}
Therefore, the longitudinal conductivity is given by
\begin{align}
	\sigma_D
	= e^2\sum_n\int \mathrm{d}k\, 
	\tau_n(\varepsilon_{nk})(v_{nk})^2
	\left(-\frac{\partial f_0(\varepsilon_{nk}-\mu)}{\partial\varepsilon_{nk}}\right)=e^2\sum_n\tau_n(E_F)v_{nF}^2g_n(E_F),
\end{align}
The current reads
\begin{align}\label{eq:j}
	j = \frac{\sigma_D}{e} \nabla \tilde{\mu} - c\frac{e}{h}(\tilde{\mu}-\tilde{\mu}_0)+j_0.
\end{align}
where the first term represents the diffusive (Ohmic) current, the second term corresponds to the chiral current, and $j_0$ is the equilibrium current.

For slowly varing electrochemical potential $\tilde{\mu}(x)$ and electric potential $U(x)$, the charge density $\rho=-e\sum_n \int \frac{\mathrm{d}k}{2\pi}f_{nk} = -e\sum_n \int \frac{\mathrm{d}k}{2\pi}f_0(\varepsilon_{nk}+U-\tilde{\mu}) \approx \rho_0-eg_0(\tilde{\mu}-\tilde{\mu}_0) + O((\tilde{\mu}-\tilde{\mu}_0)^2)$ at low temperature, where $-eg_0=\partial\rho/\partial {\tilde{\mu}}|_{\tilde{\mu}=\tilde{\mu}_0}$ and $\rho_0$ is the charge density in equilibrium.
Therefore, the Eq.~\ref{eq_ce} becomes
\begin{align}\label{eq_mu}
	-eg_0\frac{\partial \tilde{\mu}}{\partial t} =-\frac{\sigma_D}{e} \nabla^2 \tilde{\mu} + c\frac{e}{h} \nabla \tilde{\mu},
\end{align}
which is valid for $\tilde{\mu}$ slowly varying with time and smoothly varying in space. That is, the characteristic spatial and temporal scales of the inhomogeneities, $\bar{l}$ and $\bar{t}$ satisfy
\begin{align}
	\bar{l}\gg v_{nF}\tau_n(E_F), \qquad \bar{t}\gg
	\tau_n(E_F).
\end{align}
Under these conditions, the transport equation maps onto an effective Schr{\"o}dinger-like equation of motion:
\begin{eqnarray}\label{eq_rho}
i\hbar \frac{\partial \tilde{\mu}}{\partial t}  = -i[\frac{{\hat{p}^2}}{2m} + icv\hat{p}]\tilde{\mu} \equiv -iH_{\text{HN}}(\hat{p})\tilde{\mu}.
\end{eqnarray}
Here, the non-Hermitian Hamiltonian $H_{\text{HN}}(\hat{p})$ represents the continuum limit of the Hatano--Nelson (HN) model \cite{Hatano1996}, derived in the long-wavelength limit (see the next section). The operator $\hat{p}=-i\hbar\nabla$ denotes momentum, while the effective mass is given by $m=e^2 g_0\hbar/(2\sigma_D)$, and $v=1/(h g_0)$ characterizes the drift velocity. 
According to Eq.~\eqref{eq_ec}, the energy current in this model reads
\begin{align}
	j_E = \frac{\sigma_D}{e^2}\tilde{\mu}\nabla\tilde{\mu}+\frac{c}{2h}(\tilde{\mu}^2-\tilde{\mu}_0^2) + 	j_{0,E}.
\end{align}
The corresponding heat dissipation rate is given by
\begin{align}
	P_D = \nabla \cdot (j_E-j_{0,E}) =  \frac{\sigma_D}{ e^2}(\nabla\tilde{\mu})^2 + \tilde{\mu}\nabla \cdot j.
\end{align}
In the steady state, where charge conservation implies
$\nabla \cdot j = 0$, the dissipation reduces to $P_D = {\sigma_D}/{ e^2}(\nabla\tilde{\mu})^2$.

\subsection{Current expression in the presence of inelastic scattering}
Having discussed elastic scattering, we now address inelastic scattering, such as electron-phonon scattering. Equation~\ref{eq_scattering} then becomes \cite{Vasko2005}
\begin{align}\label{eq_scattering2}
	\frac{\partial f_{nk}}{\partial t} 
	+ v_{nk} \nabla f_{nk}
	- \nabla U \frac{\partial f_{nk}}{\hbar \partial k} 
	= - \sum_{n'} \int \frac{\mathrm{d}k'}{2\pi} \,
	\big[W_{nk,n'k'}  (1-f_{n'k'})f_{nk} - W_{n'k',nk}(1-f_{nk})f_{n'k'}\big].
\end{align}
Now, the scattering rate $W_{nk,n'k'}$ contains contributions from both impurity scattering and electron-phonon scattering. The detailed balance requires that $W_{nk,n'k'}/W_{n'k',nk} = e^{(\varepsilon_{nk}-\varepsilon_{n'k'})/k_B T}$.

For a non-equilibrium steady state,  
$f_{nk} = f_{nk}^{(0)} + f_{nk}^{(1)}$ with $f_{nk}^{(0)} = f_0(\varepsilon_{nk} - \mu)$.
The Boltzmann equation then reads
\begin{align}
	v_{nk} \nabla f_{nk}^{(0)} + v_{nk} \nabla f_{nk}^{(1)} 
	- \nabla U \frac{\partial f_{nk}^{(0)}}{\hbar \partial k} 
	- \nabla U \frac{\partial f_{nk}^{(1)}}{\hbar \partial k} 
	= - \sum_{n'} \int \frac{\mathrm{d}k'}{2\pi} \, \left[ W_{n'k',nk} \frac{f_{n'k'}^{(0)}}{f_{nk}^{(0)}} f_{nk}^{(1)} - W_{nk,n'k'} \frac{f_{nk}^{(0)}}{f_{n'k'}^{(0)}} f_{n'k'}^{(1)} \right],
\end{align}
where we neglect the higher-order term $f_{nk}^{(1)}f_{n'k'}^{(1)}$, which does not affect the first-order correction of $f_{nk}$.

Equivalently, introducing the matrix
\begin{equation}\label{eq_M2}
	M_{nk,n'k'} = \sum_{n''} \int \mathrm{d}k'' \, W_{n''k'',nk} \frac{f_{n''k''}^{(0)}}{f_{nk}^{(0)}} \delta_{nn'} \delta(k-k') - W_{nk,n'k'} \frac{f_{nk}^{(0)}}{f_{n'k'}^{(0)}},
\end{equation}
we obtain the same equation as Eq.~\ref{eq_27}:
\begin{align}\label{eq_36}
	- v_{nk} \frac{\partial f_{nk}^{(0)}}{\partial \varepsilon} \nabla \tilde{\mu}
	+ v_{nk} \nabla f_{nk}^{(1)} 
	- \nabla U \frac{\partial f_{nk}^{(1)}}{\hbar \partial k} 
	= - \sum_{n'} \int \frac{\mathrm{d}k'}{2\pi} \, M_{nk,n'k'} f^{(1)}_{n'k'}.
\end{align}
For inelastic collisions, the first-order correction $f_{nk}^{(1)}$ does not need to satisfy Eq.~\ref{eq_28}, but must satisfy
\begin{align}
	\sum_{n,n'} \int \frac{\mathrm{d}k}{2\pi}\,\frac{\mathrm{d}k'}{2\pi}  \, M_{nk,n'k'} f^{(1)}_{n'k'} = 0.
\end{align}
Consequently, summing over $n$ and integrating over $k$ on both sides of Eq.~\ref{eq_36} yields the current balance equation:
\begin{align}
	\sum_{n} \int \frac{\mathrm{d}k}{2\pi} \,  
	\Big( v_{nk} \nabla f_{nk}^{(1)} - \nabla U \frac{\partial f_{nk}^{(1)}}{\hbar \partial k} \Big)
	= \sum_{n} \int \frac{\mathrm{d}k}{2\pi}  \, v_{nk} 
	\frac{\partial f_{nk}^{(0)}}{\partial \varepsilon} \nabla \tilde{\mu}
	= -g_0 \, \bar{v}_0 \nabla \tilde{\mu},
\end{align}
where $g_0 = -\int \frac{\mathrm{d}k}{2\pi} \, \frac{\partial f_0(\varepsilon_{nk}-\mu)}{\partial \varepsilon_{nk}}$, and the average group velocity is defined as
\begin{align}
	\bar{v}_0
	= -\frac{1}{g_0} \sum_n \int \frac{\mathrm{d}k}{2\pi} \,
	\frac{\partial f_0(\varepsilon_{nk}-\mu)}{\partial \varepsilon_{nk}} \, v_{nk}.
\end{align}
The first-order correction $f_{nk}^{(1)}$ then satisfies
\begin{align}
	& -\frac{\partial f_{nk}^{(0)}}{\partial \varepsilon} \left( v_{nk} - \bar{v}_0 \right) \nabla \tilde{\mu}
	+ v_{nk} \nabla f_{nk}^{(1)} 
	- \nabla U \frac{\partial f_{nk}^{(1)}}{\hbar \, \partial k} \notag \\
	& \quad + \frac{\partial f_{nk}^{(0)}}{\partial \varepsilon} g_0^{-1}\sum_{n''} \int \frac{\mathrm{d}k'}{2\pi}  
	\left( v_{n''k'} \nabla f^{(1)}_{n''k'} - \nabla U \frac{\partial f^{(1)}_{n''k'}}{\hbar \, \partial k} \right)
	= -\sum_{n'} \int \frac{\mathrm{d}k'}{2\pi} \, M_{nk,n'k'} f^{(1)}_{n'k'}.
\end{align}
We retain only the first-order terms in $\nabla \tilde{\mu}$ and neglect higher-order contributions such as $\nabla^2 \tilde{\mu}$ or $(\nabla \tilde{\mu})^2$. Consequently, we obtain
\begin{align} \label{eq_f2}
	f_{nk}^{(1)} 
	= \sum_{n'} \int \frac{\mathrm{d}k'}{2\pi} \, [M^{-1}]_{nk,n'k'} 
	\left( v_{n'k'} - \bar{v}_0 \right) 
	\frac{\partial f_0(\varepsilon_{n'k'}-\mu)}{\partial \varepsilon_{n'k'}} \, \nabla \tilde{\mu},
\end{align}
which is the same as Eq.~\ref{eq_f1}. 

The electric current can be expressed as
\begin{align}
	j &= -e\sum_n \int \frac{\mathrm{d}k}{2\pi}\, v_{nk} f_{nk} \notag\\
	&= -e\sum_n \int \frac{\mathrm{d}k}{2\pi}\, \bar{v}_0 f_0(\varepsilon_{nk}-\mu)
	-e\sum_n \int \frac{\mathrm{d}k}{2\pi}\, (v_{nk}-\bar{v}_0) f_{nk}^{(1)}  \notag\\
	&= -e\sum_{n} \int \frac{\mathrm{d}k}{2\pi}\, v_{nk} [f_0(\varepsilon_{nk}-\mu)-f_0(\varepsilon_{nk}-\mu_0)] + j_0 + \frac{1}{e}\sigma_D \nabla \tilde{\mu} \notag\\
	&= \frac{1}{e}\sigma_D \nabla \tilde{\mu} - c\frac{e}{h}(\tilde{\mu}-\tilde{\mu}_0)+j_0.
\end{align}
Therefore, we obtain the same expression for the current even in the presence of inelastic scattering. Hence, all the transport results in the main text remain valid for inelastic scattering as well.
The longitudinal conductance in the presence of inelastic scattering is then given by
\begin{align}
	\sigma_D
	=  e^2\sum_{n,n'} \int \frac{\mathrm{d}k}{2\pi}\,\frac{\mathrm{d}k'}{2\pi}\,
	(v_{nk}-\bar{v}_0) [M^{-1}]_{nk,n'k'} (v_{n'k'}-\bar{v}_0)
		\left(-\frac{\partial f_0(\varepsilon_{n'k'}-\mu)}{\partial\varepsilon_{n'k'}}\right),
\end{align}
which now depends not only on impurity scattering but also on electron-phonon scattering. Note that the main effect of inelastic scattering is to renormalize the effective diffusive conductivity $\sigma_D$. 

\subsection{Equilibration mechanism}

In this subsection, we analyze how a nonequilibrium electron distribution relaxes back to equilibrium. We consider a situation where, at $t=0$, the system is driven slightly out of equilibrium by a small perturbation. The nonequilibrium distribution is written as
\[
f_{nk}(t) = f_{nk}^{(0)} + h_{nk}(t),
\]
where $f_{nk}^{(0)}$ is the equilibrium Fermi-Dirac distribution and $h_{nk}(t)$ is a small time-dependent deviation.

In the absence of external driving forces, the scattering processes described by Eq.~\ref{eq_scattering2} govern the time evolution. Substituting the above ansatz and linearizing in $h_{nk}(t)$, we obtain
\begin{align}\label{eq_ft2}
	\frac{\partial h_{nk}(t)}{\partial t} = -\sum_{n'} \int \frac{\mathrm{d}k'}{2\pi} \, M_{nk,n'k'} \, h_{n'k'}(t),
\end{align}
where the linearized collision matrix $M_{nk,n'k'}$ is given explicitly in Eq.~\ref{eq_M2}. This matrix encodes how scattering redistributes electrons among different momentum states and bands.

To reveal the relaxation dynamics more clearly, we introduce the symmetrized deviation
\[
\phi_{nk}(t) = \frac{h_{nk}(t)}{\sqrt{f_{nk}^{(0)} \bigl(1 - f_{nk}^{(0)}\bigr)}}.
\]
This transformation removes the statistical weight from the dynamics. Substituting into Eq.~\eqref{eq_ft2} yields
\begin{align}\label{eq_ft3}
	\frac{\partial \phi_{nk}(t)}{\partial t} = -\sum_{n'} \int \frac{\mathrm{d}k'}{2\pi} \, \tilde{M}_{nk,n'k'} \, \phi_{n'k'}(t),
\end{align}
where the transformed matrix is given by
\[
\tilde{M}_{nk,n'k'} = \sqrt{\frac{f_{n'k'}^{(0)}\bigl(1-f_{n'k'}^{(0)}\bigr)}{f_{nk}^{(0)}\bigl(1-f_{nk}^{(0)}\bigr)}} \; M_{nk,n'k'}.
\]

A crucial property of $\tilde{M}$ is that it is symmetric and has positive diagonal entries, as follows from the detailed balance condition. Consequently, $\tilde{M}$ is positive semidefinite, and all its eigenvalues are nonnegative. The formal solution of Eq.~\eqref{eq_ft3} is
\[
\phi_{nk}(t) = \sum_{n'} \int \frac{\mathrm{d}k'}{2\pi} \, \bigl[e^{-\tilde{M} t}\bigr]_{nk,n'k'} \, \phi_{n'k'}(0),
\]
or, equivalently,
\[
h_{nk}(t) = \sum_{n'} \int \frac{\mathrm{d}k'}{2\pi} \,
\bigl[e^{-\tilde{M} t}\bigr]_{nk,n'k'} \,
\sqrt{\frac{f_{nk}^{(0)}\bigl(1-f_{nk}^{(0)}\bigr)}{f_{n'k'}^{(0)}\bigl(1-f_{n'k'}^{(0)}\bigr)}} \; h_{n'k'}(0).
\]

Because the eigenvalues of $\tilde{M}$ are nonnegative, each mode decays exponentially in time. Therefore, $h_{nk}(t)$ relaxes to zero as $t \to \infty$, and the system returns to the equilibrium state $f_{nk}^{(0)}$.  Here, momentum relaxation is driven by static impurity scattering, whereas energy relaxation requires inelastic processes such as electron-phonon scattering.

{\bf Example: Elastic scattering and momentum relaxation.} To illustrate the general theory, we consider a concrete scenario involving only elastic perturbations. In this case, the energy of each electron is conserved during scattering. The total occupation at a given energy $\varepsilon$ remains unchanged from its equilibrium value:
\begin{align}
	\sum_{n} \int \frac{\mathrm{d}k}{2\pi} \, \delta(\varepsilon - \varepsilon_{nk}) \, f_{nk}(t)
	= \sum_{n} \int \frac{\mathrm{d}k}{2\pi} \, \delta(\varepsilon - \varepsilon_{nk}) \, f_{nk}^{(0)}.
\end{align}
Elastic scattering can redistribute momentum among electrons, thereby driving the system toward equilibrium via momentum relaxation. The scattering rate $W_{nk,n'k'}$ for this process is given by Eq.~\ref{eq_W2}, which accounts for disorder-induced transitions between states.

Solving the linearized kinetic equation, Eq.~\eqref{eq_ft2}, under the energy-conservation condition leads to explicit expressions for the relaxation of $h_{nk}(t)$. For our two-band model, we obtain
\begin{equation}
	\begin{cases}
		h_{1k}(t) = \dfrac{1}{2} \bigl[ h_{1k}(0) - h_{1,-k}(0) \bigr] 
		\exp\!\left( -\dfrac{t}{\tau_1(\varepsilon_{1k})} \right) \\[6pt]
		\qquad\quad + \dfrac{1}{2} \bigl[ h_{1k}(0) + h_{1,-k}(0) \bigr] 
		\exp\!\left( -\dfrac{g_1(\varepsilon_{1k})}{g_1(\varepsilon_{1k})+g_2(\varepsilon_{1k})} \cdot \dfrac{t}{\tau_2(\varepsilon_{1k})} \right) \\[15pt]
		h_{2k}(t) = h_{2k}(0) 
		\exp\!\left( -\dfrac{g_1(\varepsilon_{1k})}{g_1(\varepsilon_{1k})+g_2(\varepsilon_{1k})} \cdot \dfrac{t}{\tau_2(\varepsilon_{1k})} \right).
	\end{cases}
\end{equation}

These solutions show that all momentum-dependent deviations from equilibrium decay exponentially in time. The relaxation is governed by two characteristic time scales, $\tau_1(\varepsilon)$ and $\tau_2(\varepsilon)$, whose explicit forms are
\begin{equation}\label{eq_tau}
	\begin{cases}
		\dfrac{\hbar}{\tau_1(\varepsilon)} = 2\pi \left[ n_0 u_0^2 g_1(\varepsilon) + n_1 u_x^2 g_2(\varepsilon) \right], \\[10pt]
		\dfrac{\hbar}{\tau_2(\varepsilon)} = 2\pi n_1 u_x^2 \dfrac{\bigl[ g_1(\varepsilon) + g_2(\varepsilon) \bigr]^2}{g_1(\varepsilon)}.
	\end{cases}
\end{equation}

Physically, $\tau_1$ describes the relaxation of antisymmetric momentum distributions of normal mode, while $\tau_2$ controls the relaxation of chiral mode. The presence of both time scales reflects the interplay between intra-band and inter-band scattering processes. Thus, the system indeed realize momentum relaxation through elastic scattering.

\section{Non-Hermitian Continuum Model and Its Boundary Properties}

\subsection{Continuum limit of the Hatano--Nelson model}
The one-dimensional nonreciprocal Hatano--Nelson (HN) model is described by
\begin{equation}
	H_{\mathrm{HN}} = -\sum_i \left(t_R\, c_{i+1}^\dagger c_i + t_L\, c_i^\dagger c_{i+1}\right),
\end{equation}
where $t_R$ and $t_L$ are the asymmetric rightward and leftward hopping amplitudes, respectively.  
In momentum space, the Hamiltonian becomes
\begin{equation}
	H_{\mathrm{HN}}(k) = -\left(t_R e^{-ika} + t_L e^{ika}\right),
\end{equation}
yielding the complex dispersion relation
\begin{equation}
	E(k) = - (t_R + t_L) \cos(ka) + i (t_R - t_L) \sin(ka).
\end{equation}
In the long -wavelengths limit ($ka \ll 1$), the spectrum can be expanded asto second order in $k$,
\begin{equation}
	E(k) \approx -(t_R + t_L) + \frac{a^2}{2}(t_R + t_L)k^2 + i a (t_R - t_L)k.
\end{equation}
Discarding the constant energy shift, the corresponding continuum Hamiltonian reads
\begin{equation}
	H_{\mathrm{HN}} = \frac{a^2}{2\hbar^2}(t_R + t_L)\,p^2 + i\frac{a }{\hbar}(t_R - t_L)\,p.
\end{equation}
Introducing the effective parameters
\begin{equation}
	\frac{1}{2m} = \frac{a^2}{2\hbar^2}(t_R + t_L), 
	\qquad 
	v = \frac{a }{\hbar} |t_R - t_L|,
	\qquad
	c = \mathrm{sign}(t_R-t_L)
\end{equation}
the Hamiltonian can be recast in the compact form
\begin{equation}\label{eq:HNH_cont}
	H_{\mathrm{HN}} = \frac{\hat{p}^2}{2m} + icv\hat{p},
\end{equation}
which is the continuum form of the non-Hermitian Hatano--Nelson model.

\subsection{Eigenstates on the open boundary condition and skin effect}

For an open chain $x \in [-L,L]$, one can perform a  transformation
\begin{equation}
	\psi_{R}(x) = e^{c x / \lambda} \phi(x),
\end{equation}
which transforms Eq.~(\ref{eq:HNH_cont}) into the Hermitian equation
\begin{equation}
	-\frac{\hbar^2}{2m}\phi''(x) = 
	\left(E - \frac{1}{2}{m v^2}\right)\phi(x).
\end{equation}
Imposing open boundary conditions $\phi(-L)=\phi(L)=0$, the normalized solution reads
\begin{equation}
	\phi_n(x) = L^{-1/2}\sin(\kappa_n x),
\end{equation}
with eigenvalues
\begin{equation}
	E_{n,\text{open}} = \frac{\hbar^2 \kappa_n^2}{2m} + \frac{1}{2}m v^2.
\end{equation}
where $\kappa_n = n\pi/L$ with $n \in \mathbb{Z}$.
Thus the  eigenfunctions are
\begin{equation}
	\psi_{n,R}(x) = L^{-1/2} e^{c x / \lambda} \sin(\kappa_n x),
\end{equation}
which exhibit exponential localization at one boundary, manifesting the non-Hermitian skin effect.

\subsection{Green's Function of the Non-Hermitian Hamiltonian under Periodic Boundary Conditions
}

In this section, we derive the explicit form of the Green's function under periodic boundary conditions. Under periodic boundary conditions with period $2L$, the momentum is quantized as 
\[
k_n = \frac{\pi n}{L}, \qquad n \in \mathbb{Z}.
\]
The Green's function can be written in the momentum representation as
\begin{equation}
	G(E;x,x') = \bra{x}[E-H_{\text{HN}}]^{-1}\ket{x'}=\frac{1}{2L} \sum_{n \in \mathbb{Z}} \frac{e^{ik_n(x-x')}}{E - H_{\mathrm{HN}}(\hbar k_n)}.
	\label{eq:G_def}
\end{equation}
The denominator can be factorized as
\[
E - H_{\mathrm{HN}}(\hbar k) = - \frac{\hbar^2}{2m} (k - k_+)(k - k_-),
\]
where $\hbar k_\pm = -icmv \pm \sqrt{2mE-(cmv)^2}$ 
are two complex roots satisfying the equation $E = H_{\mathrm{HN}}(\hbar k)$. Substituting this form into Eq.~(\ref{eq:G_def}) yields
\begin{equation}
	G(E;x,x') = -\frac{m}{\hbar^2 L} \frac{1}{k_+ - k_-}
	\sum_{n \in \mathbb{Z}} e^{ik_n(x-x')}
	\left( \frac{1}{k_n - k_+} - \frac{1}{k_n - k_-} \right).
	\label{eq:G_sum}
\end{equation}
The discrete summation in Eq.~(\ref{eq:G_sum}) can be evaluated by the contour-integration identity
\[
\sum_{n \in \mathbb{Z}} \frac{1}{2L} \frac{e^{ik_n s}}{k_n - k_0}
= i\,\frac{e^{ik_0 s}}{1 - e^{i2L k_0}}, \qquad 0 \leq s < 2L,
\]
where $s = x - x'$. Using this result, we obtain
\begin{eqnarray}
	G(E;x,x') 
	&=& -\frac{2i m}{\hbar^2 (k_+ - k_-)}
	\left[
	\frac{e^{ik_+(x-x')}}{1 - e^{i2L k_+}}
	- \frac{e^{ik_-(x-x')}}{1 - e^{i2L k_-}}
	\right].
\end{eqnarray}
By defining $\kappa = (k_+ - k_-)/2 = \sqrt{2mE-(cmv)^2}/\hbar$ and including the non-Hermitian correction factor $e^{c(x-x')/\lambda}$ with $\lambda = \hbar/(mv)$, 
the Green function takes the compact form
\begin{eqnarray}
	G(E;x,x') = -\frac{i m}{\hbar^2 \kappa}
	\left(
	\frac{e^{i \kappa (x-x')}}{1 - e^{i 2L k_+}}
	- \frac{e^{-i \kappa (x-x')}}{1 - e^{i 2L k_-}}
	\right) e^{\frac{c(x-x')}{\lambda}},
	\label{eq:G_periodic}
\end{eqnarray}
which is valid for $0 < x - x' < 2L$. For other cases, the Green's function satisfies the periodic condition
\[
G(E;x,x' \pm 2L) = G(E;x,x').
\]
\begin{figure}[bht]
	\centering
	\includegraphics[width=5in]{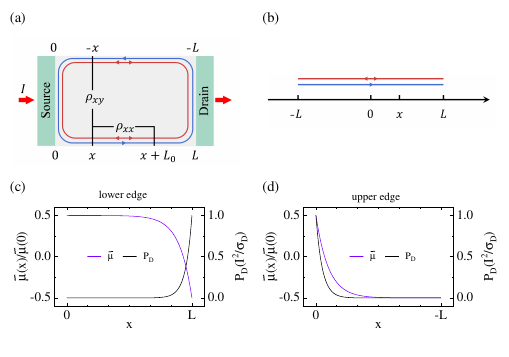}
	\caption{(Color online). 
		(a) Schematic of a Hall-bar device hosting one chiral and one normal mode. The positive directions of the local 1D boundary coordinates for the upper and lower edges follow their respective chiral propagation directions.
		(b) One-dimensional coordinate describing the spatial distribution of the edge states in (a) along the boundary.
		(c,d) Spatial distributions of the electrochemical potential $\tilde{\mu}(x)$ and heat dissipation $P_D$ 
		for the lower and upper edges, respectively, with $c=1$.  
		All parameters are the same as in Fig.~2 of the main text.		
		\label{figs1}}
\end{figure}

\subsection{Spatial Profiles of the Electrochemical Potential}
We consider a setup with a source at $x=0$ and a drain at $x=L$ [see Fig.~\ref{figs1}], described by the source term $S(x) = I\delta(x) - I\delta(x-L)$.
Solving the charge continuity equation (Eq.~\ref{eq_ce2}) for the steady-state  electrochemical potential $\tilde{\mu}(x)$ leads to the relation:
\begin{eqnarray}\label{eq_V}
	\tilde{\mu}(x) = \frac{\hbar I}{eg_0}[G(0;x,0)-G(0;x,L)]
\end{eqnarray}
This equation directly links the experimentally measurable electrochemical potential (or voltage profile, $V(x)=-\tilde{\mu}(x)/e$) to the Green's function of the non-Hermitian Hamiltonian. 
Using the Green's function
\begin{eqnarray}
	G(0;x,x') = -\frac{m\lambda}{\hbar^2 |c|}
	\left(
	\frac{e^{c_+ (x-x')/\lambda}}{1 - e^{c_+ 2L /\lambda}}
	- \frac{e^{c_- (x-x')/\lambda}}{1 - e^{c_- 2L /\lambda}}
	\right),
	\quad \text{for } 0 < x - x' < 2L,
\end{eqnarray}
where $c_{\pm}=c\pm|c|$, we obtain
\begin{equation}\label{eq_s}
	\tilde{\mu}(x) =
	\begin{cases}
		\dfrac{Ih}{ec}
		\left(
		-\dfrac{1}{2}+
		\dfrac{e^{2c x/\lambda}}{1 + e^{-2cL /\lambda}}
		\right),
		& -L \leq x \leq 0, \\[10pt]
		\dfrac{Ih}{ec}
		\left(
		\dfrac{1}{2}-\dfrac{e^{2c x/\lambda}}{1 + e^{2cL /\lambda}}
		\right),
		& 0 \leq x \leq L .
	\end{cases}
\end{equation}
The first line describes the upper edge, and the second line describes the lower edge.
As shown in Fig.~\ref{figs1}(c) and (d) for $c=1$, both the electrochemical potential $\tilde{\mu}$ and heat dissipation $P_D$ exhibit pronounced exponential spatial profiles on the two edges. However, the spatial localization occurs at opposite ends of the Hall bar: for the lower edge, $\tilde{\mu}$ and $P_D$ are localized near $x=L$ (the drain terminal), whereas for the upper edge, they are localized near $x=0$ (the source terminal).

\section{Comparison between numerical results and theoretical predictions}
\begin{figure}[bht]
	\centering
	\includegraphics[width=7in]{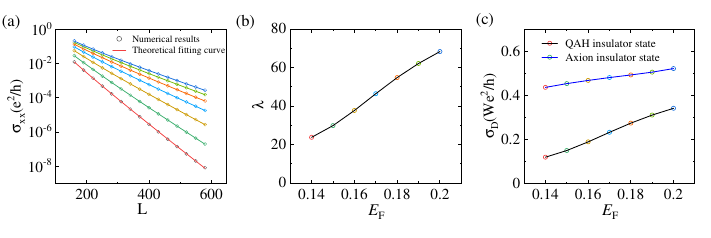}
	\caption{(Color online). 
		(a) Longitudinal conductivity $\sigma_{xx}$ as a function of system size $L$ on a logarithmic scale in the QAH insulator state. Open circles represent numerical results, while solid lines denote theoretical fitting curves. Different colors correspond to different Fermi energies shown in (b).
	    (b) The characteristic length $\lambda$ extracted from the fitting curve as a function of Fermi energy $E_F$.
		(c) One-dimensional conductivity $\sigma_D$ as a function of Fermi energy $E_F$. The upper (lower) set of data corresponds to the QAH insulator (axion insulator) state. Other parameters are the same as in Fig.~3 of the main text.
		\label{figs2}}
\end{figure}

In this section, we compare the numerical results with the theoretical results. Based on the theoretical derivation, the longitudinal conductivity is expressed as
\begin{equation}\label{eq_s}
	\sigma_{xx}(L) =
	|c|\,\dfrac{e^2}{h}\,
	\dfrac{W}{L_0}
	\Bigl[
	\Theta(c)\, e^{2c(x+L_0 - L)/\lambda}
	+\Theta(-c)\, e^{2c x/\lambda}
	\Bigr],
	\qquad \text{for } x,\, L \gg \lambda .
\end{equation}
In our numerical calculation model, we set $L_0 = L/3$ and $x = L/3$. Equation~\ref{eq_s} can then be simplified to
\begin{equation}\label{eq_s2}
	\sigma_{xx}(L) = \sigma_{xx}(L_1)\,\frac{L_1}{L}\,
	e^{-\frac{2|c|(L-L_1)}{3\lambda}},
	\qquad \text{for } L \gg \lambda .
\end{equation}
We use this theoretical expression of the longitudinal conductivity to fit the numerical results, as shown in Fig.~\ref{figs2}(a). The fitting curve based on Eq.~\ref{eq_s2} agrees very well with the numerical data, demonstrating consistency between theory and numerical calculations.

From the fitting, we further extract the characteristic length $\lambda$ (see Fig.~\ref{figs2}(b)) and the corresponding diffusive conductivity $\sigma_D = \lambda e^2/2h$ (see Fig.~\ref{figs2}(c)). 
The extracted $\sigma_D(c=1)$ in the QAH state increases with the Fermi energy $E_F$, consistent with the enhanced density of normal edge states at higher energies.
Moreover, $\sigma_D(c=1)$ 
in the QAH state is of the same order as that in the axion-insulator state, as expected because both are governed by diffusive normal edge modes. Their quantitative difference arises from the different normal-mode dispersions in the two magnetic configurations, as well as from the presence or absence of the chiral edge channel.


Therefore, the B{\" u}ttiker virtual leads should not be viewed as an additional mechanism unrelated to the kinetic theory. Rather, they provide a numerical implementation of the dephasing and local-equilibrium conditions required for the continuum transport description. The quantitative agreement between the analytical fits and the numerical simulations supports the validity of our quantum kinetic theory and the experimental feasibility of observing the predicted non-Hermitian transport response in magnetic topological insulators.
\bibliographystyle{apsrev4-2}
\bibliography{suppref.bib}